\documentclass{nature}
\usepackage{graphicx} 
\usepackage{color}
\usepackage{xcolor}
\usepackage{caption}

\usepackage[normalem]{ulem} 
\usepackage{amsmath}
\usepackage{amssymb}
\usepackage{float}
\usepackage{rotating}

\usepackage[super,comma,sort&compress]{natbib}
\makeatletter 
\let\saved@includegraphics\includegraphics 
\AtBeginDocument{\let\includegraphics\saved@includegraphics} 
\renewenvironment*{figure}{\@float{figure}}{\end@float} 
\makeatother 

\begin{document}

\title{Observation of suppressed viscosity in the normal state of $^3$He due to superfluid fluctuations }

\author{Rakin N. Baten$^1$, Yefan Tian$^1$, Eric N. Smith$^1$, Erich J. Mueller$^1$, Jeevak M. Parpia$^{1*}$}

\maketitle
\begin{affiliations}
    \item[$^1$]{Department of Physics, Cornell University, Ithaca, N.Y. 14853, USA}
    
    \item[*]{jmp9@cornell.edu}
\end{affiliations}

\date{\today}

\begin{abstract}
\begin{center}
    Abstract:
\end{center}
By monitoring the quality factor of a quartz tuning fork oscillator we have observed a 
fluctuation-driven reduction in the viscosity of bulk $^3$He in the normal state near the superfluid transition temperature, $T_c$. These fluctuations, which are only found within $100 \mu$K of $T_c$, play a vital role in the
theoretical modeling of ordering; they encode details about the Fermi liquid parameters, pairing symmetry, and scattering phase shifts.  They will be of crucial importance for transport probes of the topologically nontrivial features  of superfluid $^3$He under strong confinement.  Here we characterize the temperature and pressure dependence of the fluctuation signature, finding 
data collapse consistent with the predicted theoretical behavior.
\end{abstract}

\flushbottom
\maketitle
%
%
\thispagestyle{empty}
\newpage

\section*{Introduction}
The normal state of a superfluid contains transient ordered patches which grow as the system is cooled towards the transition temperature $T_c$.  Observing the influence of these fluctuations on transport in liquid $^3$He has been a scientific goal which has been unfulfilled for nearly 50 years \cite{EmeryJLTP76}. Similar fluctuations are found near other ordered states, such as magnets\cite{andersonmagnet}, superconductors\cite{Aslamazov1996}, and alkali gases\cite{Randeria2014}, where they are often related to pseudogap phemomena\cite{Mueller2017,Timusk1999}. 
These fluctuations have been particularly well studied in $^4$He, where the extremely short coherence length allows the $\lambda$ anomaly in the heat capacity of $^4$He
to serve as a model system for scaling\cite{Chui1996}.
Due to the low pairing energy and long coherence length, finding such signatures in $^3$He, however, has been challenging.  Here we observe a fluctuation induced suppression of the viscosity of bulk $^3$He near $T_c$.  This provides crucial information about the transport signature which can be used to probe contemporary phenomena such as the topologically nontrivial nature of superfluidity in confined $^3$He\cite{Mizushima2015}.

The low temperature normal state of $^3$He is our best example of a  Fermi liquid, whose properties are understood in terms of a gas of interacting quasiparticles\cite{Abrikosov1959}.  As the temperature is lowered, the phase space available for scattering is reduced and the mean time between scattering events grows as $\tau\propto T^{-2}$.  As a consequence, transverse momentum gradients produce smaller stresses at low temperatures, quantified by the viscosity $\eta\propto\tau\propto T^{-2}$.  A scattering resonance emerges as the liquid is cooled towards the superfluid transition, where particles form short-lived Cooper pairs during scattering events. Such resonances enhance the scattering, leading to a decrease in the viscosity. In a clean 3D system (such as $^3$He), this suppression occurs in only a very narrow window of temperature $\delta T=T-T_c$ where the pair lifetime $\tau_{\rm GL}\approx \hbar/k_B \delta T$ is comparable to $\tau$.  Thus one only expects to see a measurable reduction of the viscosity at temperatures of order $1\%$ above $T_c$.  In principle the nature of these fluctuations will change when one is within the scaling regime\cite{Ginzburg1950,Larkin2008} $\delta T/T_c =(T_c/T_F)^4\approx 10^{-12}$, but in practice such precision is unachievable. 

In addition to being of fundamental interest, the fluctuation contributions to transport are important for future experiments which will look for edge modes\cite{SaulsPRB2011,WuandSauls2013,Saulsarxiv2022} in $^3$He as  a signature of topological superfluidity\cite{Levitin13Science,Levitin13PRL,Zhelev17NC,Heikkinen2021,LotnykNatComm2020,LotnykPRL2021}.  The contribution to viscosity from these edge modes will be small, and accurate measurements will be needed to distinguish them from the effects of fluctuations.  Here we report the necessary base-line measurements.

Fluctuation effects in $^3$He have previously been observed in the attenuation of zero (collisionless) sound 
\cite{PaulsonPRL78,Samalam1978,McClintock1978}, with ever increasing experimental and theoretical sophistication \cite{Lee1996HighFA,Granroth1998BroadbandFS,Pal1979FluctuationCT,Sauls2022}.  While valuable, these are not a substitute for transport experiments.   
Observing the fluctuation contributions to viscosity is challenging and  previous attempts\cite{ParpiaPRL78,Tian2021,Carless1983,Nakagawa1996} have had flaws which obscured or complicated the phenomena.  In this work we overcome these challenges.

Firstly, Refs. \citen{Carless1983,Nakagawa1996} observed significant deviation from Fermi-liquid behavior ($\eta\propto T^{-2}$) at all temperatures.  Such deviations are unphysical, and are not seen in heat capacity\cite{Greywall86SH}, thermal conductivity\cite{Greywall84TC}, in collisonless sound measurements \cite{PaulsonPRL78}, or in previous measurements with quartz forks\cite{Blaauwgeers07}.  The deviations may be due to the temperature dependence of the properties of the metallic alloys used as vibrating elements\cite{Morley2002}.  We avoid this issue by using quartz forks.

\begin{figure}
 \renewcommand{\figurename}{Figure}
\centering
\includegraphics[width=0.85\textwidth]{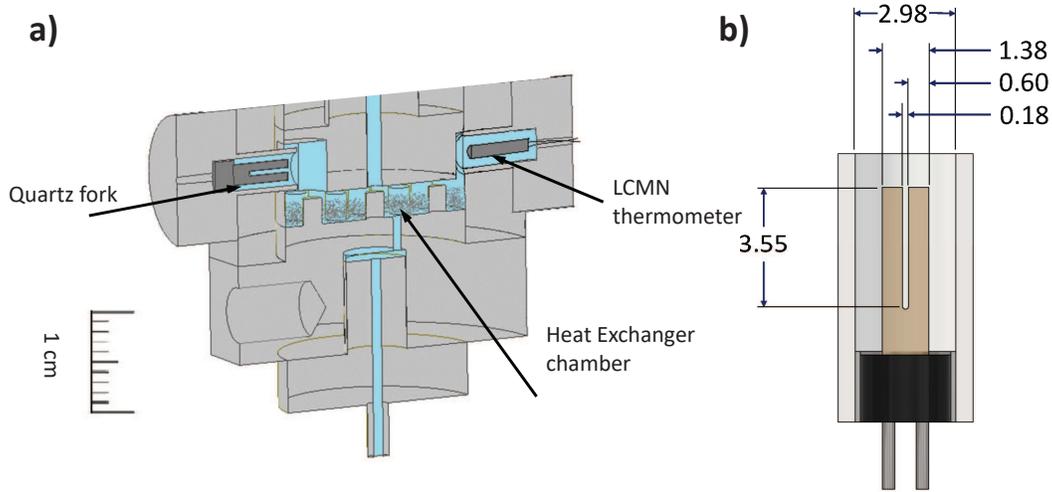}
\caption{\small
  {\bf Cell Schematic} a) The location of the quartz fork and LCMN thermometer are shown in relation to the heat exchanger. 
  b) Schematic image of the quartz fork with dimensions in millimeters.}
   \label{fig::1CellandFork}
   
\end{figure}

Secondly, Refs. \citen{ParpiaPRL78,Tian2021} inferred temperature from the susceptability of a small sample of undiluted Cerous Magnesium Nitrate (CMN).  While accurate at $\sim10$mK, this approach suffers from systematic errors near the magnetic ordering temperature of CMN. Our current experiment uses a Lanthanum Diluted Cerous Magnesium Nitrate thermometer (Figure ~\ref{fig::1CellandFork}a), referencing thermometry to the widely accepted PLTS2000 temperature scale\cite{PLTS2000,Tian2022}.

Finally, our experiment takes pains to work within the hydrodynamic regime, where the viscous mean free path $\lambda_\eta$ is small compared to all other relevant length-scales.  In Refs.~\citen{ParpiaPRL78,Tian2021}, $\lambda_\eta$ was comparable to the cavity height at low pressure, leading to slip, and deviations from Fermi-liquid behavior which obscured the influence of fluctuations.  Torsional oscillator experiments\cite{ParpiaPRL1983} find that the contributions from these Knudsen effects become observable when the device dimensions are $d\approx 8\lambda_\eta$.
In the present work, our fork has  tines which are 0.61 mm wide $\times$ 0.253 mm thick $\times$ 3.64 mm long, spaced 0.194 mm apart,  housed in a cylindrical casing $\approx$3 mm in diameter (Figure ~\ref{fig::1CellandFork}b).  The smallest of these dimensions, the 0.194 mm tine spacing, is more than 8 times $\lambda_\eta$ except at the very lowest temperatures (see Supplemental Note 1, Supplemental Table 1). Thus, Knudsen effects should be negligible.

\flushbottom
%
%
\thispagestyle{empty}

\section*{Results}

We monitor the quality factor $Q= f_0/\Delta f$ of a quartz fork\cite{Blaauwgeers07} immersed in liquid $^3$He cooled to mK temperatures by a nuclear demagnetization stage\cite{Parpia84RSI}.  Here, $f_0$ is the resonant frequency and $\Delta f$ is the resonance linewidth.  The oscillator damping can be related to the helium viscosity ($Q\propto \eta^{-1/2}$)\cite{Blaauwgeers07}, and we operate in 
the hydrodynamic regime.  Temperature was measured with a diluted paramagnetic salt thermometer placed in the same $^3$He volume proximate to the quartz fork. Additional details on thermometry, fork operation, Fermi liquid viscosity, the hydrodynamic regime and background subtraction are provided in the methods section and in Supplementary Notes 1 and 2. 
The pressure was maintained at a constant value using electronic feedback for each temperature sweep.

\begin{figure}
 \renewcommand{\figurename}{Figure}
\centering
\includegraphics[width=0.80\textwidth]{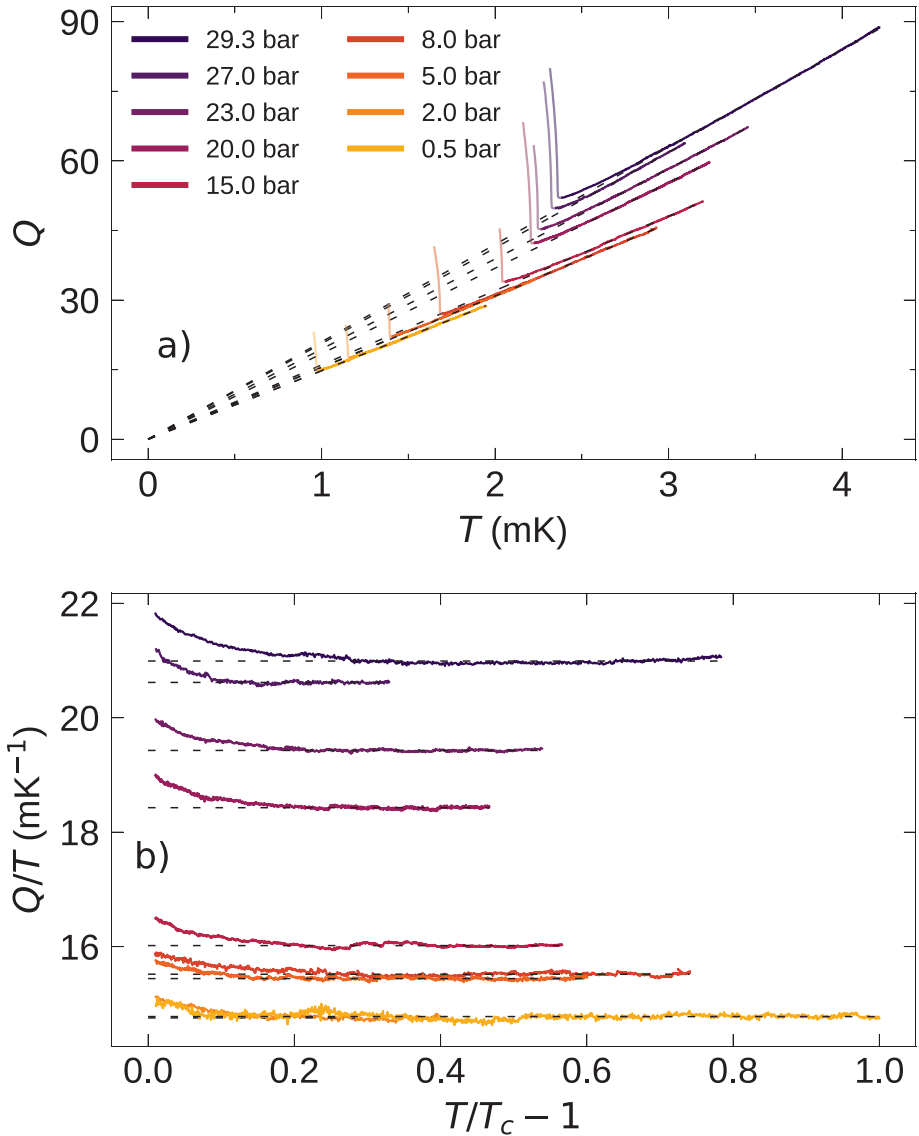}

\caption{
{\bf Quartz Fork $Q$ vs Temperature} a) The inferred $Q$ of the quartz fork at various pressures {\it vs} the temperature. The expected Fermi liquid behavior was obtained after processing described in Supplementary Note 2, ($Q \propto \eta^{-1/2} \propto T$), and is represented by dashed lines. The superfluid transition is marked by an abrupt increase in the $Q$, and data below $T_c$ is shown as lighter shaded lines. b) Plot of $Q/T$ vs ($1-T/T_c$). This plot illustrates the extent of the departure of $Q$ from linear behavior  with pressure. } 
   \label{fig::2QvsT}

\end{figure}

\begin{figure}
 \renewcommand{\figurename}{Figure}
\centering
\includegraphics[width=0.8\textwidth]{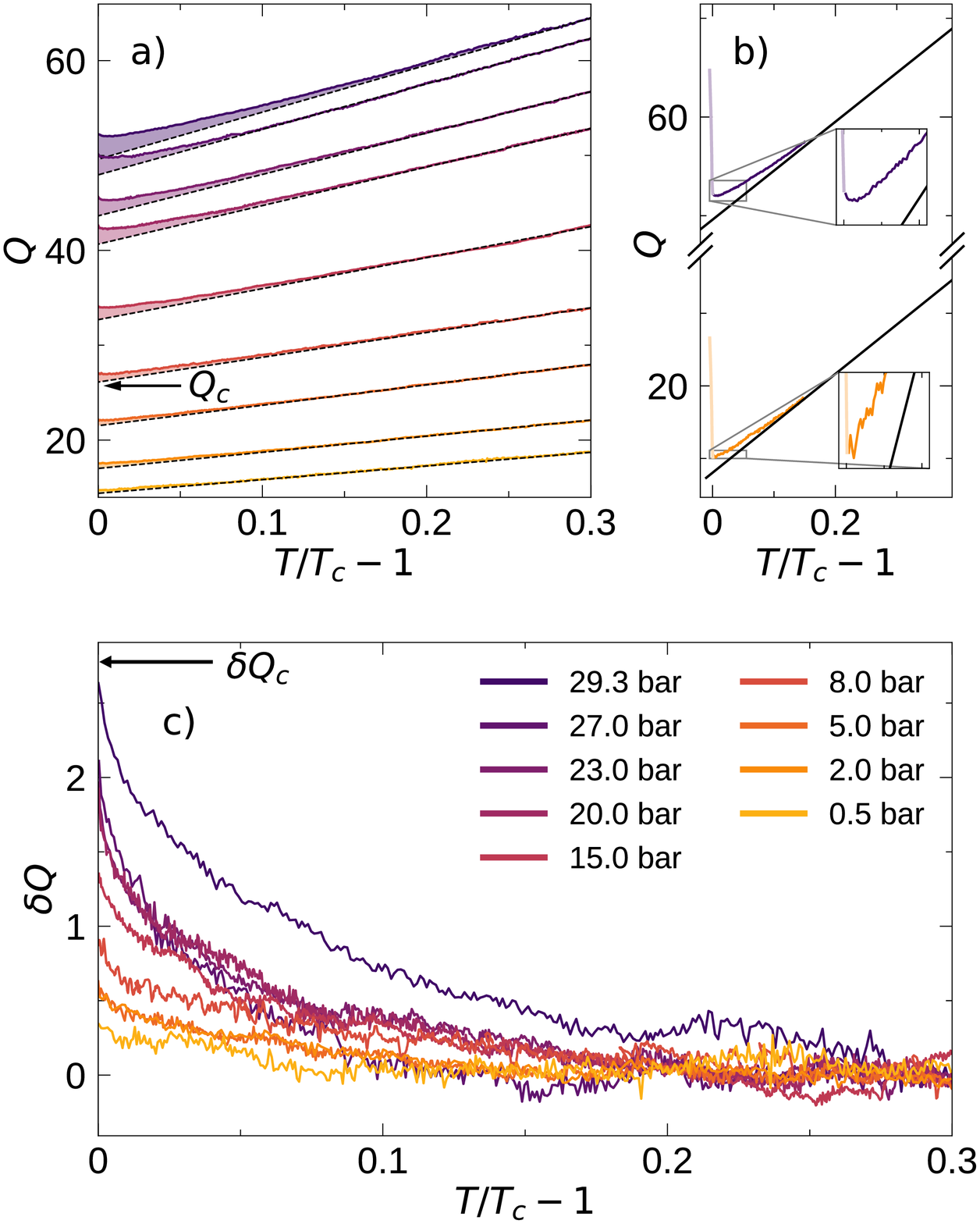}

\caption{\small
  {\bf Quartz Fork $Q$ vs Temperature near $T_c$}  a) Departure from the Fermi liquid behavior (linear slope) is seen at all pressures just above $T_c$. Also marked is the value of $Q_c$ for the 8 bar run. b)  $Q$ {\it vs} ($T/T_c - 1$) near $T_c$ of the 29.3 bar and 2 bar runs. It is evident from the insets that the higher pressure run shows a minimum in the $Q$ before $T_c$ is attained, while the lower pressure data shows no minimum. c)  The excess $Q$  $vs$ $T/T_c -1$. 
   \label{fig::3QvsT}
  }  
\end{figure}

The data obtained at several pressures from 0.5 bar to 29.3 bar are shown in Figure~\ref{fig::2QvsT} a). For each data set we show the best linear fit as a dashed line passing through the origin, corresponding to the Fermi liquid prediction $\eta\propto T^{-2}$ (ie. $Q\propto T$). In Figure ~\ref{fig::2QvsT} b), we compare the value of $Q/T$ obtained at all pressures near $T_c$, illustrating the extent of the departure from Fermi liquid behavior near $T_c$.

As $T_c$ is approached from above (Figure~\ref{fig::3QvsT}a)), a small increase in $Q$ ($ \delta Q$) is observed relative to the dashed line, corresponding to a suppression of $\eta$. At high pressure, the deviations are large enough that $Q$ actually passes through a minimum in the normal state. At low pressure, $\delta Q$ is smaller, though it can be resolved. The differences between high and low pressure results are highlighted in Figure~\ref{fig::3QvsT} b) and its insets. Upon entering in to the superfluid state the $Q$ sharply increases due to the rapid decrease in viscosity\cite{Alvesalo1973, ParpiaPRL78,BhattacharyyaPRL1975,BhattacharyyaPRB1977} at $T_c$. The quality of the data is sufficient to illustrate the development of $\delta Q$  in Figure \ref{fig::3QvsT} c) with  pressure.

\section*{Discussion}

Proximity to superfluidity enhances quasiparticle scattering:  Quasiparticles that pass near each other form short-lived pairs, increasing the scattering rate, $1/\tau$. 
The viscosity is proportional to the scattering time $\tau$, ($\propto T^{-2}$), which is therefore suppressed near $T_c$. Emery\cite{EmeryJLTP76} writes the fluctuation contribution to the viscous scattering time $\tau$ as 

\begin{equation}\label{emery}
\frac{\delta{\tau}}{\tau}= -\Gamma \left(\frac{k_BT_F\tau}{\hbar}\right)(k_F\xi_{00})^{-3}\alpha \left(1-\frac{\theta^{1/2}}{\alpha} {\mathrm{tan}}^{-1}\frac{\alpha}{\theta^{1/2}}\right) 
\end{equation}

\noindent where the quantity $\delta \tau$ is the additional scattering time due to the broken pairs above $T_c$, and $\alpha$ is a fitting constant. Here $\theta = \frac{T}{T_c}-1$ is the reduced temperature, $T_F$ is the Fermi temperature, and $\Gamma$ is a numerical constant that depends on the pairing and the transport parameter (in this case viscosity, $\eta$). 
The unitless quantity $k_F \xi_{00}$ is the product of the Fermi wavevector and the pairing coherence length, and in bulk $^3$He can be expressed as 
\begin{equation}\label{emery2}
\left( k_F \xi_{00}\right) ^2 = \frac{7\zeta(3)}{12\pi^2}\left(\frac{T_F}{T_c}\right)^2
\end{equation}
\noindent where $\zeta(3)\approx1.2$ is Ap\'ery's constant and $\zeta$ is the Riemann Zeta function.

\begin{figure}
 \renewcommand{\figurename}{Figure}
\centering
\includegraphics[width=0.85\textwidth]{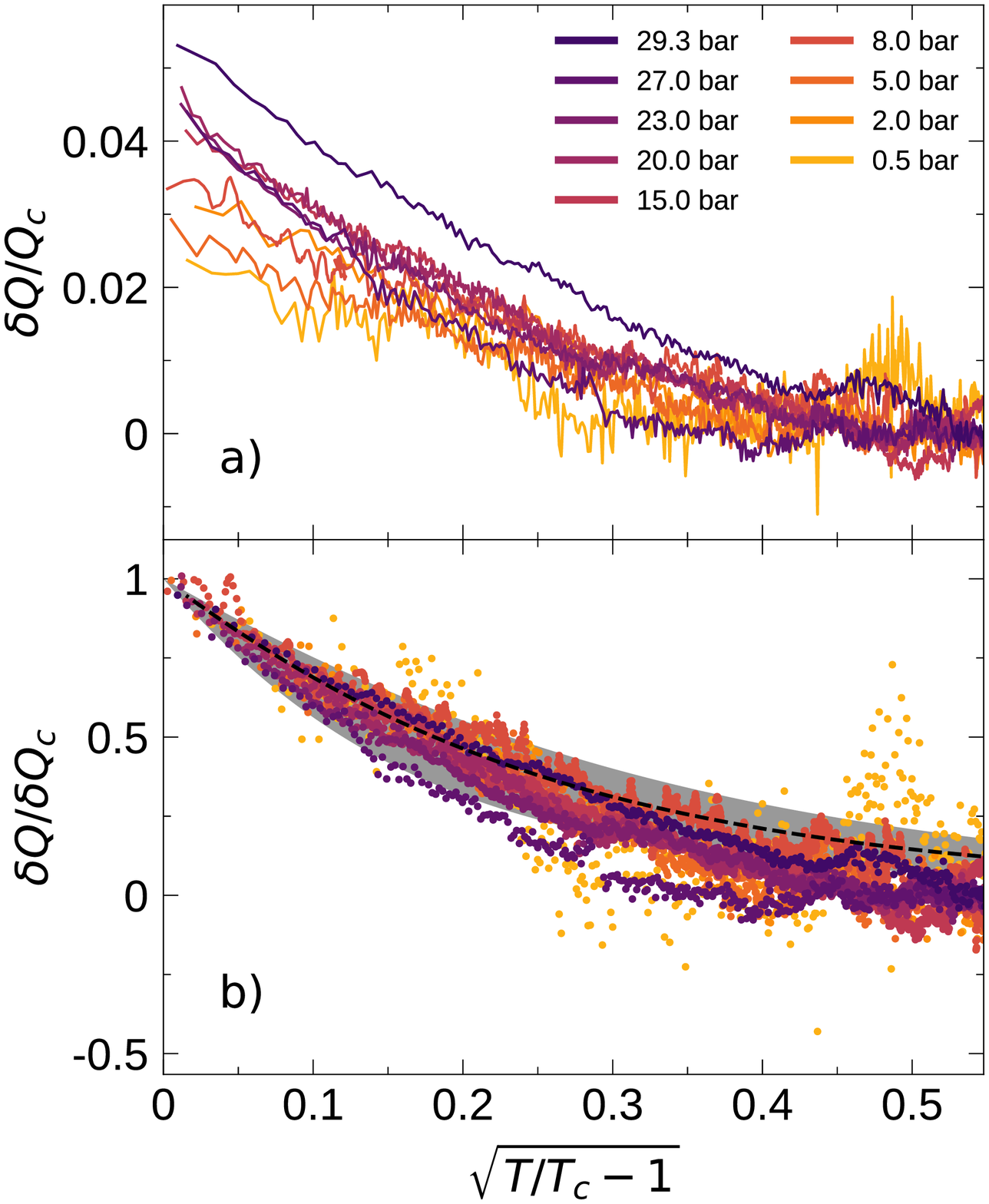}

\caption{\small
  {\bf Normalized fluctuation contribution {\it vs} reduced temperature} a) The measured excess $Q$ (see Figure \ref{fig::3QvsT}) of the quartz fork at various pressures, normalized to $Q_c$ (see {Figure \ref{fig::3QvsT} c)}) plotted against the square root of reduced temperature. This plot shows that the contribution to $Q$ of the fluctuation component increases faster than the increase of $Q_c$ with pressure. b) $\delta Q$ normalized to $\delta Q_c$ (see Eqs. \ref{emery3},~\ref{coef3}). 
   The dashed line shows the expected temperature dependent fit to the fluctuation component of viscosity in Eq. (\ref{emery}) (see  Ref~[\citen{EmeryJLTP76}]).
The shaded gray region represents the 1$\sigma$ range in the curve fit based upon the error in the fit parameters.
  \label{fig::4QnearTc}  
  } 
\end{figure}

Since the $Q \propto \eta^{-1/2} \propto \tau^{-1/2}$, it follows that $\delta Q/Q =  - 1/2 ~\delta \tau/\tau$. We can rewrite $\tau(T)$ = $\tau_c \times (T_c/T)^2$ and  $Q(T)$= $Q_c \times T/T_c$. $Q_c$ is the value of the $Q$ at $T_c$ without the contribution due to fluctuations (See Figure~\ref{fig::3QvsT}b)). Thus ($\delta \tau/\tau)$= - 2 ($\delta Q/Q_c$) $\times$ ($T_c/T$). 
This yields a modified version of Equation (\ref{emery}),

\begin{equation}\label{emery3}
\begin{split}
\frac{\delta Q(T)}{Q_c} 
&= C(P)\frac{\alpha}{1+\theta}\left(1-\frac{\theta^{1/2}}{\alpha} {\mathrm{tan}}^{-1}\frac{\alpha}{\theta^{1/2}}\right),
\end{split}
\end{equation}
and
\begin{equation}\label{coef3}
C(P)=\frac{1}{2}\Gamma \left(\frac{k_BT_F\tau_c}{\hbar}\right)(k_F\xi_{00})^{-3}.
\end{equation}

We can extract $Q_c$ from the linear fits in Figure~\ref{fig::3QvsT} and plot the ratio $\delta Q/Q_c$ from Eq.~(\ref{emery3}) in Figure~\ref{fig::4QnearTc} a). 
For small $\theta$, Eq.~(\ref{emery3}) has the form $\delta Q\approx\delta Q_c(1-\pi \theta^{1/2}/2\alpha)$, where $\delta Q_c$ is the excess $Q$ at $T_c$.  Thus, it is natural to use $\theta^{1/2}$ as the horizontal axis.
  Both $Q_c$ and $\delta Q$ increase with pressure, but $\delta Q$ has a slightly stronger dependence:  The ratio $\delta Q_c/Q_c$ varies from $\sim$2\% at the lowest pressure measured to $\sim$5\% at the highest. The corresponding values of the zero sound attenuation coefficient, $\mathrm{A}$, $\delta\mathrm{A}/\mathrm{A}_c$ measured in collisionless sound varied from $\sim$8\% at 32.56 bar, $\sim$6.5\% at 19.94 bar and ``very approximately $\sim$2\%" at 0.05 bar\cite{PaulsonPRL78}. Assuming that $\alpha$ is not pressure dependent, Eq.~(\ref{emery3}) predicts that the excess $Q$'s should collapse if normalized  as $\delta Q/\delta Q_c$.  In Figure \ref{fig::4QnearTc} b) we test that feature, showing Emery's prediction as a black dashed line, using  $\alpha=0.43$.  The agreement is quite remarkable, with slight deviations at larger values of $\theta^{1/2}$.

\begin{figure}
\renewcommand{\figurename}{Figure}
\centering
\includegraphics[width=0.8\textwidth]{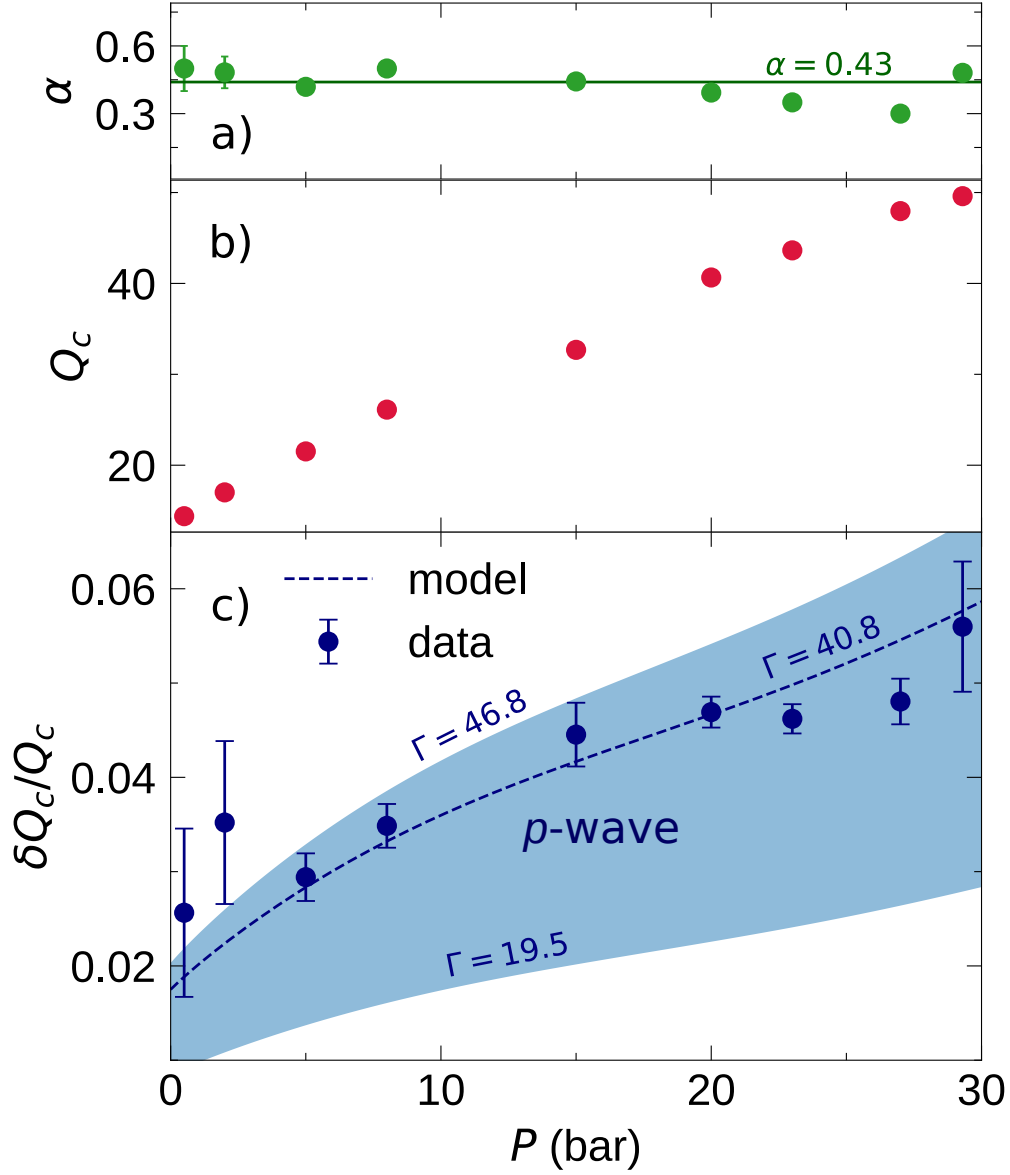}
\caption{\small
  {\bf Comparison of measured and calculated parameters} 
    {\bf a)} The values of $\alpha$ obtained to yield the fit shown in Figure \ref{fig::4QnearTc} b). The horizontal line denotes the mean $\alpha$. 
    {\bf b)} Values of $Q_c$ based on the linear fits for Fermi liquid behavior shown in Figure~\ref{fig::3QvsT}.
    {\bf c)} $\delta Q_c$, normalized by $Q_c$ plotted against the pressure. The dashed line shows the expected temperature dependent fit to the normalized fluctuation component in Eq. \ref{emery3} (see  Ref~[\citen{EmeryJLTP76}]) based on previously measured values of $\eta T^2$ (Ref~[\citen{ParpiaThesis}]), $v_F$, $n$ and $m^*/m$ (Ref~[\citen{Greywall86SH}, \citen{Greywall83PRB}]). Details in Supplementary Note 1, Supplementary Table 1. Shaded blue region marks the variation of $\Gamma$ for $p$ wave pairing in the model by Emery (Ref~[\citen{EmeryJLTP76}]). 
    \label{fig::3Comparison}} 
  \label{fig::5Comparison}  
  
\end{figure}

We further quantify this agreement by independently fitting each fixed-pressure run to Eq.~(\ref{emery3}), extracting our best estimates of the pressure dependence of $\delta Q_c=C(P) \alpha Q_c$ and $\alpha$.  As seen in Figure \ref{fig::5Comparison} a), any pressure dependence of $\alpha$ is weak.  The contributions to $C(P)$ in Eq. (\ref{coef3}) are reasonably well known.  We take
$\eta T^2$  from Ref~[\citen{ParpiaThesis}] to calculate $\tau_c$ (after correction for temperature scales), and $v_F$, $n$ and $m^*/m$ from Ref~[\citen{Greywall86SH}, \citen{Greywall83PRB}]; (See Supplementary Note 1 for more details).  Emery argues that $19.5<\Gamma<46.8$ for $p$ wave pairing, with the true value likely lying in the middle of that range.
We treat $\Gamma$ as a free parameter,  finding a best-fit value $\Gamma=40.8$, which is at the upper end of the expected range.  Nonetheless, the resulting curve, shown in Fig.~\ref{fig::5Comparison} b), agrees very well with our measurements.  The error bars on $\alpha$, $\delta Q_c/Q$ in Figure~\ref{fig::5Comparison} a), b) represent a 1$\sigma$ standard deviation. The error bars on $\alpha$ are derived from the calculation of the fit to Equation~\ref{emery3} and random noise error in $Q$; the error bars on $\delta Q_c/Q_c$ in  Figure~\ref{fig::5Comparison} b) are derived from the error in $\delta Q_c$ (the error in $Q_c$ is negligible in comparison to $\delta Q_c$).

The somewhat large value of $\Gamma$ may be the result of limitations in Emery's modeling.  Lin and Sauls\cite{Sauls2022} argued that Emery's calculation contains some double-counting, and that it incorrectly included interference terms among the different scattering channels.  Another source of theoretical uncertainty is the scattering time $\tau$ which we used in evaluating Eq.~(\ref{coef3}).  In any event, the magnitudes of the fluctuation contribution to the viscosity are seen to be smaller than the values noted in References [\citen{PaulsonPRL78,Samalam1978}].

With improvements in signal recovery using low temperature amplifiers, the precision and noise of the excess $Q$ could be greatly improved, and perhaps used to measure the pressure dependence of the Landau parameter $F_2^s$ as was proposed for collisionless sound\cite{Sauls2022}.
The values of $F_2^s$ are poorly known\cite{Sauls2022}, as they are derived from the pressure dependence of the attenuation of transverse zero sound which a difficult to measure parameter\cite{KettersonPRL1976}.  

Looking forward, an important next step will be to extend these measurements to strongly confined geometries, where topological surface states appear\cite{SaulsPRB2011,WuandSauls2013,Saulsarxiv2022,Levitin13Science,Levitin13PRL,Zhelev17NC,Heikkinen2021,LotnykNatComm2020,LotnykPRL2021}.  In such geometries $T_c$ can be significantly suppressed\cite{Heikkinen2021}, leaving an extended region where fluctuations can potentially become stronger.
Experiments studying thermal transport in such narrow channels \cite{LotnykNatComm2020} reveal a crossover between bulk and surface dominated regimes, which depend on surface quality \cite{Sharma2018,Autti2020,Heikkinen2021}.  The role of pairing fluctuations, and their interaction with surface modes, has not yet been established, and will be the focus of future research. For the present study conducted in bulk $^3$He, the impact of surface states (that exist only below $T_c$) on fluctuations should be negligible.


We have observed that incipient pairing fluctuations contribute a small but significant portion of the scattering above $T_c$. This contribution is resolved at all pressures, and is comparable to that observed using the attenuation of collisionless (zero) sound. There are significant efforts underway to study transport processes such as mass and spin edge currents\cite{SaulsPRB2011,WuandSauls2013,SaulsPRB2013}, thermal Hall effects \cite{Saulsarxiv2022}, thermal conductivity\cite{LotnykNatComm2020} and spin diffusion in highly confined geometries, where the suppression of $T_c$ and strong confinement should lead to the enhancement of the contribution of fluctuations, potentially impacting exotic topological transport.

\section*{Methods}

{\bf Quartz fork:} The experimental results described here were obtained with a quartz fork\cite{Blaauwgeers07} with dimensions much greater than the quasiparticle mean free path. The other relevant length scale is the viscous penetration depth, $\delta = (2\eta/\rho\omega)^{1/2}$, where $\eta$ and $\rho$ are the viscosity and density of the $^3$He, while $\omega$ is the resonant frequency of the fork. The largest value of the viscous penetration depth occurs at $T_c$ at 0 bar. Unlike collisionless sound where $\omega\tau\geq$1, here the fork operates in the hydrodynamic limit ($\omega\tau\le$1) with $\omega$ = 2$\pi f_0\approx$ 2$\times$10$^5$ $s^{-1}$ and $\tau \approx$ 2 $\times$10$^{-6} s$ at $p = $~0 bar and $T=T_c$ (see Supplementary Note 1 for further details).

\noindent{\bf Fork operation:} The quartz fork was operated in a phase locked loop and driven at a fixed drive voltage. The phase locked loop was set to drive the fork at a frequency fixed to within 5 Hz from resonance. When the frequency shift exceeded these bounds, the drive frequency was adjusted to bring the device on resonance again. The resonant frequency and $Q$ were inferred from the complex response recorded by the lock in amplifier. In order to simplify this conversion, a significant background response of the non-resonant signal (``feedthrough") had to be measured and subtracted from the received signal. After subtraction, when the drive frequency was swept through resonance, the signal was seen to be Lorentzian, and was calibrated to yield the $Q$. Further details are provided in the Supplementary Note 2. 

\noindent{{\bf Thermometry:} Thermometry was accomplished using a small pill (1.25 mm diameter, 1.25 mm high) of $\le$30 $\mu$m diameter powdered Lanthanum diluted Cerous Magnesium Nitrate (LCMN), packed to 50\% density. The pill and monitoring coil were located in a niobium shielding can. The coil structure consisted of an astatically wound secondary and primary coil. The primary coil was driven at constant voltage through a 10 k$\Omega$ resistor by a signal generator at fixed frequency (23Hz). The secondary coil was coupled to the input of a SQUID. The secondary loop had an additional mutual inductor to allow cancellation of the induced signal in the loop. The input of this mutual inductor was driven by the same signal generator as the primary. The drive amplitude and phase of this cancellation signal was stepped by discrete amounts to cancel out most of the current in the secondary loop. The drive applied to the mutual inductor and the magnitude of the received signal were proportional to the susceptibility of the LCMN. These were calibrated against a melting curve thermometer and against the superfluid transition temperatures at various pressures. The thermometer had a resolution of better than 50 $nK$. }



\begin{addendum}
\item[Acknowledgements]

This work was supported by the National Science Foundation, under DMR-2002692 (JMP), and PHY-2110250 (EJM). 

\item[Author contributions statement]

Experimental work was principally carried out by Y.T. and R.B. with further support from E.N.S. and J.M.P.  Analysis and the presentation of figures was carried out by R.B. and Y.T.. We thank Anna Eyal for generating the figure of the cell in Figure 1. E.M significantly contributed to the analysis and the writing of the manuscript, J.M.P. supervised the work and J.M.P., and E.M. had leading roles in formulating the research and writing this paper. R.B and Y.T contributed equally to the publication of this result. All authors contributed to revisions to the paper.
\item[Correspondence] All correspondence should be directed to jmp9@cornell.edu
 \item[Data Availability] The  data generated in this study and shown in all the plots in this paper and the supplementary material have been deposited in the Cornell
University e-commons data repository database under accession
code https://doi.org/10.7298/r4jy-py94.

\end{addendum}

\section*{Additional information}

\textbf{Competing Interests:}
The authors declare that they have no competing interests.

\newpage
\makeatletter

\renewcommand \thesection {Supplementary Note \@arabic\c@section:}

\makeatother

\setcounter{figure}{0}
\renewcommand*{\citenumfont}[1]{S#1}
\renewcommand*{\bibnumfmt}[1]{[S#1]}

\section{Calculation of $\tau_{\eta} T^2$ and other parameters}

 \begin{figure}[H]
 \renewcommand{\figurename}{Supplementary Figure}
\centering
\includegraphics[width=0.7\linewidth, keepaspectratio]{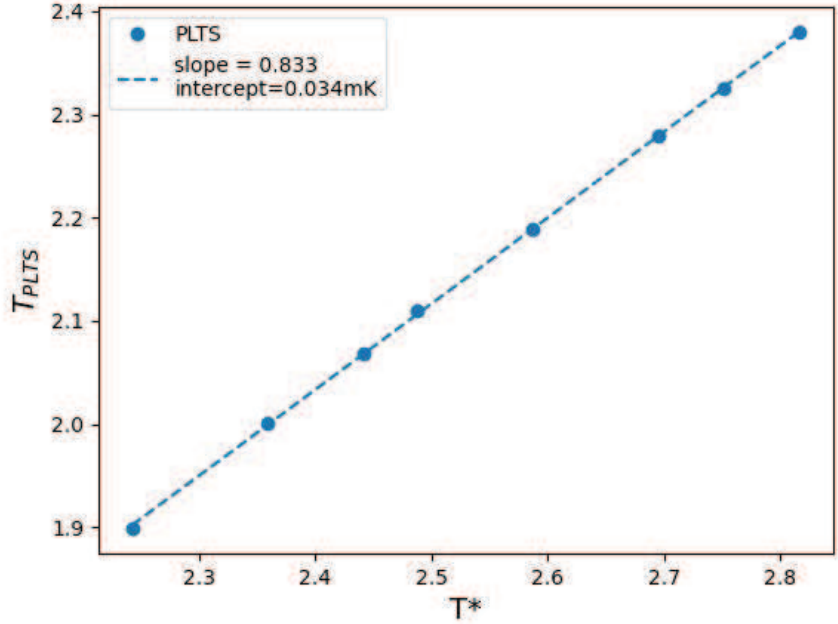}
\caption{{\bf{Conversion from $T^*$ to $T_{PLTS}$}}. Values of $T_c^*$ [\citen{ParpiaLT15, ParpiaThesis}] plotted against values of $T_c$ on the $PLTS$ scale\cite{PLTS2000,PLTS}. }
\label{fig::S1}
\end{figure}

To arrive at an estimate for the magnitude of the fluctuation contribution to the $Q$, we need to determine various pressure dependent quantities for liquid $^3$He. We start with the determination of $\tau_\eta T^2$, the quasiparticle scattering time associated with the viscosity. The Fermi liquid viscosity was studied by Parpia and co-workers\cite{ParpiaLT15, ParpiaThesis}. The temperature scale used in that work needs to be converted to the PLTS scale\cite{PLTS2000}. We plot the values of $T_c^*$  against the values of $T_c$ in the PLTS scale\cite{PLTS} in Supplementary Figure~\ref{fig::S1}. The conversion requires a linear scaling with a small offset, yielding $T_{PLTS} = 0.833T^* + 0.034$. The pressure dependent viscosity coefficients $\eta T^2$ [poise-mK$^2$] listed in Ref[\citen{ParpiaLT15}] are then converted to their values with the PLTS scale.

 \begin{figure}[H]
 \renewcommand{\figurename}{Supplementary Figure}
\centering
\includegraphics[width=0.7\linewidth, keepaspectratio]{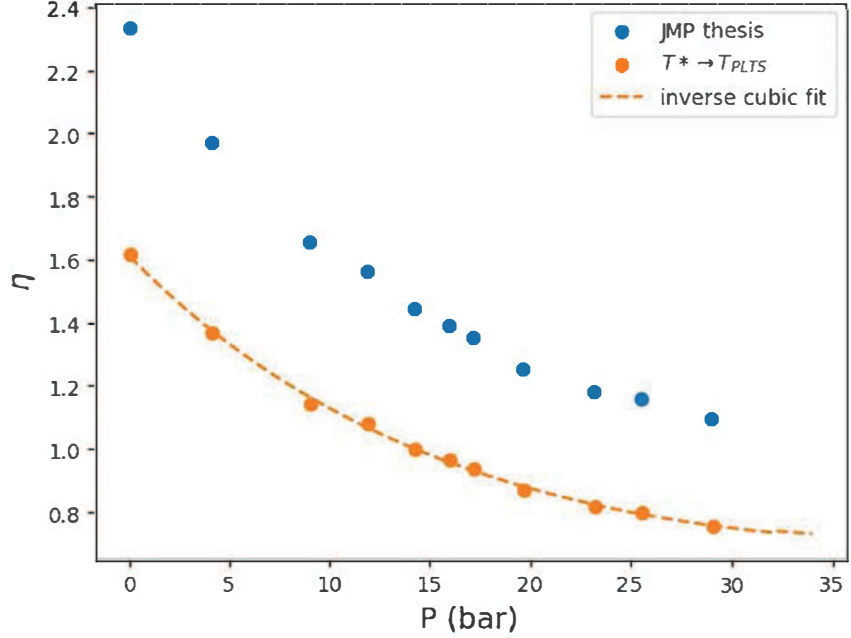}
\caption{{\bf{$\eta T^2$ vs Pressure}}. Values of the viscosity coefficient\cite{ParpiaLT15} before conversion (blue) and after conversion to the PLTS scale (gold). The dashed line is a cubic fit to $\eta (P) T^2$. }
\label{fig::S2}
\end{figure}

The viscosity coefficient, $\eta (P) T^2$ ($\eta$ in poise, $P$ in bar, $T$ in mK  following the PLTS scale), shown in Supplementary Figure \ref{fig::S2} and listed in Supplementary Table 1, can be calculated from the relation, 

\begin{equation}
  \begin{aligned}
   & (\eta(P)T^2)^{-1}=\sum_{i=0}^3 A_i P^i, \\
   &\rm{with} \\
   & A_0=6.18470415\times10^{-1}, && A_1=2.49235869\times10^{-2}, \\
   & A_2=2.33758602\times10^{-4}, && A_3=-9.37151796\times10^{-6}. \\
  \end{aligned}
  \label{Eq::etaTsq}
\end{equation}

The viscosity of $^3$He can be written as 

\begin{equation}
  \begin{aligned}
   & \eta =\frac{1}{5}np_F\lambda = \frac{1}{5}nm\left(\frac{m*}{m}\right) v_F^2\tau,\\
   &\rm{thus} \\
& \tau(P)T^2=5\eta(P)T^2\frac{V_m}{5.009\times10^{-24}\times N_A} \frac{1}{(m^*/m)v_F^2},
    \end{aligned}
  \label{Eq::eta}
\end{equation}

\noindent with $p_F$ and $v_F$ the Fermi momentum and velocity respectively, $n$, the particle density, $\lambda$ the quasiparticle mean free path, $\tau$ the viscous scattering time, $V_m$ the molar volume, $N_A$, Avogadro's number, and $m$*/$m$ the effective mass ratio. The mass of each $^3$He atom, $m$ = 5.009 $\times$10$^{-24}$ g, is also needed to obtain the mass density. 
 
The molar volume (in cm$^3$, with $P$ in bar) is reproduced from Reference [\citen{Greywall86SH}].

\begin{equation}
  \begin{aligned}
   & V_{\rm{m}}=\sum_{i=0}^5 B_i P^i, \\
   &\rm{with} \\
   & B_0=36.837231, && B_1=-1.1803474, \\
   & B_2=8.3421417\times10^{-2}, && B_3=-3.8859562\times10^{-3}, \\
   & B_4=9.475978\times10^{-5}, && B_5=-9.1253577\times10^{-7}. \\
  \end{aligned}
  \label{Eq::MolarVolume}
\end{equation}

\noindent The effective mass was obtained by fitting a polynomial to the data in Reference [\citen{Greywall86SH}].

\begin{equation}
  \begin{aligned}
   & \frac{m^*}{m}=\sum_{i=0}^4C_i P^i, \\
   &\rm{with} \\
   & C_0=2.80, && C_1=0.1292, \\
   & C_2=-3.188\times10^{-3}, && C_3=9.372\times10^{-5}, \\
   & C_4=-1.03\times10^{-6}. \\
  \end{aligned}
  \label{Eq::mstarm}
\end{equation}

\noindent The Fermi velocity was obtained by fitting a polynomial to the data in Reference [\citen{Greywall86SH}]

\begin{equation}
  \begin{aligned}
   & v_F=\sum_{i=0}^4D_i P^i, \\
   &\rm{with} \\
   & D_0=59.8353052512899, && D_1=-1.99579000023344, \\
   & D_2=8.19564928730876\times10^{-2}, && D_3=-2.10299440496278\times10^{-3}, \\
   & D_4=2.13615981987922\times10^{-5}. \\
  \end{aligned}
  \label{Eq::vF}
\end{equation}

\noindent A polynomial fit to $T_c(P)$ on the PLTS scale is provided in Reference [\citen{PLTS}] and listed here 
\begin{equation}
  \begin{aligned}
   & T_{\rm{c,PLTS}}=\sum_{i=0}^5 E_i P^i, \\
   &\rm{with} \\
   & E_0=0.90972399274531, && E_1=0.14037182852625, \\
   & E_2=-0.0074017331747577, && E_3=2.8617547367067\times10^{-4}, \\
   & E_4=-6.5064429600510\times10^{-6}, && E_5=6.0754459040296\times10^{-8}. \\
  \end{aligned}
  \label{Eq::TcPLTS}
\end{equation}

We use Supplementary Equations \ref{Eq::etaTsq}, \ref{Eq::eta}, \ref{Eq::MolarVolume}, \ref{Eq::mstarm}, \ref{Eq::vF} to calculate $\tau_{\eta} T^2$ and then Supplementary Equation \ref{Eq::eta}, \ref{Eq::TcPLTS} to calculate the quasiparticle scattering time at $T_c$, $\tau_c$. We also need $T_F$ to calculate the quantity $C(P)$ in Equation 4 of the main paper. We use Equation 6 in Reference [\citen{Greywall83PRB}] together with Supplemental Equation \ref{Eq::mstarm} to calculate $T_F$ (K), using $V_m$ in cc/mole.

\begin{equation}
  \begin{aligned}
   & T_{F}=\frac{\hbar^3}{2mk_B(m^*/m)}\left(\frac{3\pi^2}{V_m}\right)^{2/3} = \frac{54.91}{m^*/m}V_m^{-2/3}. \\
  \end{aligned}
  \label{Eq::Tf}
\end{equation}

\noindent Values for the coefficient $\tau_{\eta} T^2$,  $\tau_c$, $T_c$, $V_m$, $v_F$, $T_F$ and $m^*/m$ at various pressures are listed in Supplementary Table 1. These values can be used along with the best fit determination of $\alpha$ = 0.434 and $\Gamma$ = 40.8 to obtain values for C(P) in Equation 3 of the main paper, which also relates to $\delta Q_c/Q_c$, the maximum contribution to the excess $Q$ at the transition temperature plotted in Figure 3 a) of the main paper. 

The hydrodynamic regime ($\omega\tau \le$ 1), is distinct from the collisionless regime ($\omega\tau \ge$ 1).  In the hydrodynamic regime, collisions are frequent compared to the frequency of the external excitation ({\it e.g.} pressure oscillation or shear frequency). In the collisionless regime, the excitation frequency exceeds the inverse mean time between collisions. In our experiment, the largest value of $\omega\tau_c$ is attained for P = 0 and $T_c$ and is calculated to be  $\approx$ 0.243 (see Supplementary Table \ref{emerytable}). In the first observation of collisionless sound\cite{Abel1966}, the attenuation of first sound is $\approx$ 4\% below the expected normal state value for $\omega\tau \approx 0.25$. Thus, it is possible that at the lowest pressure, a portion of the deviation from $Q \propto T$ is due to a deviation from hydrodynamic behavior. However, such a contribution would be distributed over a range of temperature and not confined to the region near $T_c$. Additionally, the increase in $T_c$ with pressure and the decrease of the coefficient $\eta T^2$ (and $\tau T^2$) with pressure assures us that the departure from Fermi-liquid behavior cannot be accounted for by non-equilibrium effects due to any departure from the ideal $\omega\tau \ll 1$ regime. 

The mean free path ($\lambda_\eta$) should be smaller than the viscous penetration depth $\delta$=$(2\eta/\rho\omega)$, which is the distance over which the transverse velocity field in a fluid of density $\rho$, viscosity $\eta$, in contact with an object oscillating at a frequency $\omega$, decays exponentially. The mean free path should also be smaller than the confinement size. These conditions are met well at high pressure, and marginally at low pressure (see Supplementary Table \ref{emerytable}). Once again, the observation of the departure from Fermi-liquid behavior is strongest at high pressure, where the conditions for hydrodynamic behavior are well fulfilled. 

\begin{sidewaystable}

\small
\centering 
\begin{tabular}{|l|l|l|l|l|l|l|l|l|l|l|l|}
\hline

Pressure [bar] & 0& 3 & 6& 9&12&15&18&21&24&27&30 \\
\hline
$\tau_{\eta} T^2$ [$\mu s mK^2$] & 0.985 & 0.861 & 0.776 & 0.712 & 0.660 & 0.616  & 0.597 & 0.550 & 0.529 & 0.515 & 0.508\\
$\tau_c$ [$\mu s$] & 1.19 & 0.533 & 0.328 & 0.234 & 0.182 & 0.149  & 0.127 & 0.111 & 0.100 & 0.0933 & 0.0887\\
\hline
$T_c$ [$mK$]  & 0.9097 & 1.271 & 1.539 & 1.743 & 1.903 & 2.033 & 2.139 & 2.226 & 2.296 & 2.351 & 2.392  \\
\hline
$T_F$ [K] & 1.77 & 1.66 & 1.56  & 1.49 & 1.42 & 1.36 & 1.31 & 1.26 & 1.22 & 1.17 & 1.13\\
\hline
$V_m$ [$cm^3$] & 36.84 & 33.95 & 32.03 & 30.71 & 29.71 & 28.89 & 28.18 & 27.55 & 27.01 & 26.56 & 26.17 \\
\hline
$m^*/m$  & 2.80 & 3.16 & 3.48 & 3.77 & 4.03 & 4.28 & 4.53 & 4.77 & 5.02 & 5.26 & 5.50 \\
\hline
$v_F$ [$m/s$]  & 59.83 & 54.53 & 50.38 & 47.11 & 44.49 & 42.32 & 40.44 & 38.74 & 37.15 & 35.65 & 34.24 \\
\hline
$\delta Q/Q_c$ & 0.0176   & 0.0244    & 0.0302    &0.0343     &  0.0382    & 0.0416    & 0.0445   & 0.0474    & 0.0500     & 0.0544   & 0.0582   \\
\hline
$\omega\tau_c$ & 0.243 & 0.109 & 0.0670 & 0.0478 & 0.0372 & 0.0304 & 0.0259 & 0.0227 &0.0204 &0.0191 &0.0181 \\
\hline
$\lambda_{\eta}(T_c)$ [$\mu m]$& 71.2&29.1&16.5&11.0&8.10&6.31&5.14&4.30&3.72&3.33&3.04 \\
\hline
$\lambda_{\eta}($\rm{ 3}~$ mK)$ [$\mu m]$ & 6.55&5.22&4.35&3.72&3.26&2.90&2.61&2.37&2.18&2.04&1.93 \\
\hline
$\eta (T_c)$ [poise] & 1.97&0.890&0.544&0.384&0.296&0.239&0.202&0.174&0.155&0.141&0.132 \\
\hline
$\delta_c$ [$\mu m]$ & 154&99.2&75.3&61.9&53.5&47.4&43.0&39.6&36.9&35.0&33.5 \\
\hline
\end{tabular}
\caption{\bf{Supplementary Table 1: Fermi liquid parameters}} Lists various quantities to estimate the fluctuation contribution in Equation 3, Main article, and for the discussion concerning mean free paths.  
\label{emerytable}
\end{sidewaystable}

\pagebreak
\pagebreak

\section{Background correction procedure}

\section*{Outline of procedure}
In this Supplementary Note, we describe details of the procedure we followed to correctly subtract the non-resonant background signal from our quartz tuning fork. The quartz fork is driven and detected through its in-built piezo electric capability. The distinguishing feature of our data on $Q$ is the fact that we are able to track the quartz fork's $Q$ continuously at all pressures. This is enabled by our use of a digital phase locked loop (PLL) that keeps the quartz fork on or near resonance. There is significant electrical coupling of the drive signal to the output side, with an attendant frequency dependent phase shift. The loop requires that the received vector signal $X(T), Y(T)$ from the lock in amplifier, has the non-resonant signal (feed through) subtracted from the received signal (See Supplementary Figure~\ref{fig::S3}). When operated in vacuum, the fork's $Q$ is high enough so that the received signal displays a Lorentzian response without background subtraction. When operated in liquid $^3$He at low temperatures, the fork's $Q$ can be as low as $\approx$10 leading to a broad response with a correspondingly small resonant signal requiring subtraction of the background signal for further analysis. 

The subtraction procedure was carried out while gathering the data within the LabView Virtual Instrument environment. However, after the data was accumulated at several pressures, it became apparent that the original background calibration was insufficiently precise and that a post processing procedure would have to be followed. If the fitted background was used ``as is", the result would be a non-Lorentzian resonance seen in the red trace in Supplementary Figure~\ref{fig::S5}. (The background subtraction and effects are shown in Supplementary Figures~\ref{fig::S3},~\ref{fig::S4},~\ref{fig::S5}). An important condition imposed was that the $Q$ {\it vs} $T$ plot have a $Q(T = 0) =0$ conforming with Fermi liquid behavior expectations. During our temperature sweep we adjusted the drive frequency when it differed from the inferred resonant frequency by 5 Hz.   If the background subtraction is not performed correctly one finds small jumps in the inferred $Q$ and resonant frequency.  We used the elimination of these jumps as an additional constraint. 

This Supplementary note details the procedure for all post-acquisition adjustments at one pressure (29.3 bar). Briefly, the procedure consisted of

\begin{enumerate}
    \item A linear fit to the ``as collected" $Q$ {\it vs} $T$ for temperatures between $1.2T_c$ and $1.4T_c$ yielded the
  intercept, $Q(0)_1$, listed as the first iteration of the correction procedure in Supplementary Table ~\ref{tab:Qadjtable}.  $Q(0)_1$ was converted to signal units (Volts) using a constant $k$ (defined later), and was subtracted from the $X(T)$ data, shown in Supplementary Figure ~\ref{fig::S6}. This ``enforced" expectations of Fermi liquid behavior in the normal liquid ($\eta \rightarrow \infty$ as $T \rightarrow 0$).
    \item The Nyquist trace $Y(T)$ {\it vs} $X(T)$ was plotted for each constant pressure cooldown. Circular arcs of constant $Q$ were drawn for each drive-frequency reset, shown in Supplementary Figure ~\ref{fig::S7}a). The $X(T)$ data segments for a given fixed drive frequency were shifted to the nearest constant $Q$ arc, shown in Supplementary Figure ~\ref{fig::S7}b). This was done to eliminate the slightly jagged character of the $Q$ {\it vs} $T$ plot. The mean displacement between $X(T)$ and its nearest constant $Q$ arc (denoted as $\overline{\Delta Q_1}$) is the mean jump in $Q$ found in the ``as collected" data and is tabulated in Supplementary Table ~\ref{tab:Qadjtable}. $Q(T)$ was recalculated with the corrected $X(T)$ data and the ``as collected" $Y(T)$ data. This completes the first iteration of $X$ offset corrections.
    \item A linear fit to the recalculated $Q(T)$ data between $1.2T_c$ and $2T_c$ (or the highest temperature in the run), yielded $Q(0)_2$, the intercept in the second iteration of the $X$ offset correction procedure (tabulated in Supplementary Table~\ref{tab:Qadjtable}). $Q(0)_2$ was converted to Volts with the $k$ constant, and was subtracted from the $X(T)$ data. The Nyquist trace was plotted, and circular arcs of constant $Q$ were calculated for each frequency reset point. $X(T)$ data segments at fixed drive frequency were shifted onto the nearest constant $Q$ arc. The mean displacement between $X(T)$ and its nearest constant $Q$ arc, $\overline{\Delta Q_2}$, the mean jump in $Q$ found in the second iteration of $X$ offset corrections, is tabulated in Supplementary Table ~\ref{tab:Qadjtable}. The $Q(T)$ is recalculated with the corrected $X(T)$ data and the ``as collected" $Y(T)$ data.  
    This completes the second iteration of $X$ offset corrections.
    \item  A linear fit to the doubly recalculated $Q(T)$ data for temperatures between $1.2T_c$ and $2T_c$ (or the highest temperature in the run), yields $Q(0)_3$, the intercept in the second iteration of the $X$ offset correction procedure tabulated in Supplementary Table ~\ref{tab:Qadjtable}. $Q(0)_3$ was converted to Volts with the $k$ constant, and was subtracted from the $X(T)$ data. The Nyquist trace was plotted, and the circular constant $Q$ arcs were calculated and drawn at each frequency reset point. $X(T)$ data segments for a fixed drive frequency were shifted onto the nearest constant $Q$ arc. The mean displacement between $X(T)$ and its nearest constant $Q$ arc, $\overline{\Delta Q_3}$, (the mean jump in Q found in the third iteration of $X$ offset corrections) is tabulated in Supplementary Table ~\ref{tab:Qadjtable}. The $Q(T)$ was recalculated with the corrected $X(T)$ data and the ``as collected" $Y(T)$ data.
    This completes the third iteration of $X$ offset corrections.
    \item The resonant frequency, $f_o(T)$, was recalculated after the three iterations of $X$ offset corrections. It displayed discontinuous line segments, with jumps in $f_o(T) $ at each frequency reset point. (See Supplementary Figure~\ref{fig::S8}). The sum of the jumps at each $f_o$ was minimized by multiplying the ``originally set" $k$ in LabView by a multiplicative constant, $k_{adj}$, listed Supplementary Table~\ref{tab:Qadjtable}. The final $Q(T)$ was recalculated with the new scaling constant $k_{adj} \times k$. A linear fit to the $Q(T)$ for temperatures between $1.2T_c$ and $2T_c$ (or the highest temperature in the run) was obtained, and its slope $T_c$ is the $Q_c$ reported in the main body of the paper. The intercept of this line is the final intercept quoted in Supplementary Table~\ref{tab:Qadjtable}. Supplementary Figure~\ref{fig::S9} compares the final recalculation of $Q(T)$ and its linear fit, with the ``as collected" $Q(T)$ and the linear fit calculated in step 1.
\end{enumerate}

In the following, we provide more details accompanied by figures to clarify the procedure. Importantly, the fluctuation precursor is seen in the ``raw data" before the various iterations at all pressures. The post data-acquisition procedure is needed to provide the ``Fermi liquid background" behavior to scale the fluctuation contribution. We list the procedure and details so that other users may adapt it for their own investigations.  Elimination or significant reduction of the non-resonant background signal is essential to resolve any finer detailed variation of the fluctuation contribution. Ultimately, the ability to continuously track the $Q$ together with the high resolution thermometry enabled the fluctuation contribution to the viscosity of $^3$He to be resolved in this experiment. 

\section*{Background subtraction and first iteration.}

As stated in the summary, we recorded values of $X(T), Y(T)$ obtained while driving a quartz resonator at a frequency, $f_D$ near the fork's resonant frequency, $f_0$ immersed in $^3$He. To calculate the $Q$ and $f_0$, the non-resonant signal has to be first subtracted from the received signal.

 \begin{figure}[H]
 \renewcommand{\figurename}{Supplementary Figure}
\centering
\includegraphics[width=\linewidth, keepaspectratio]{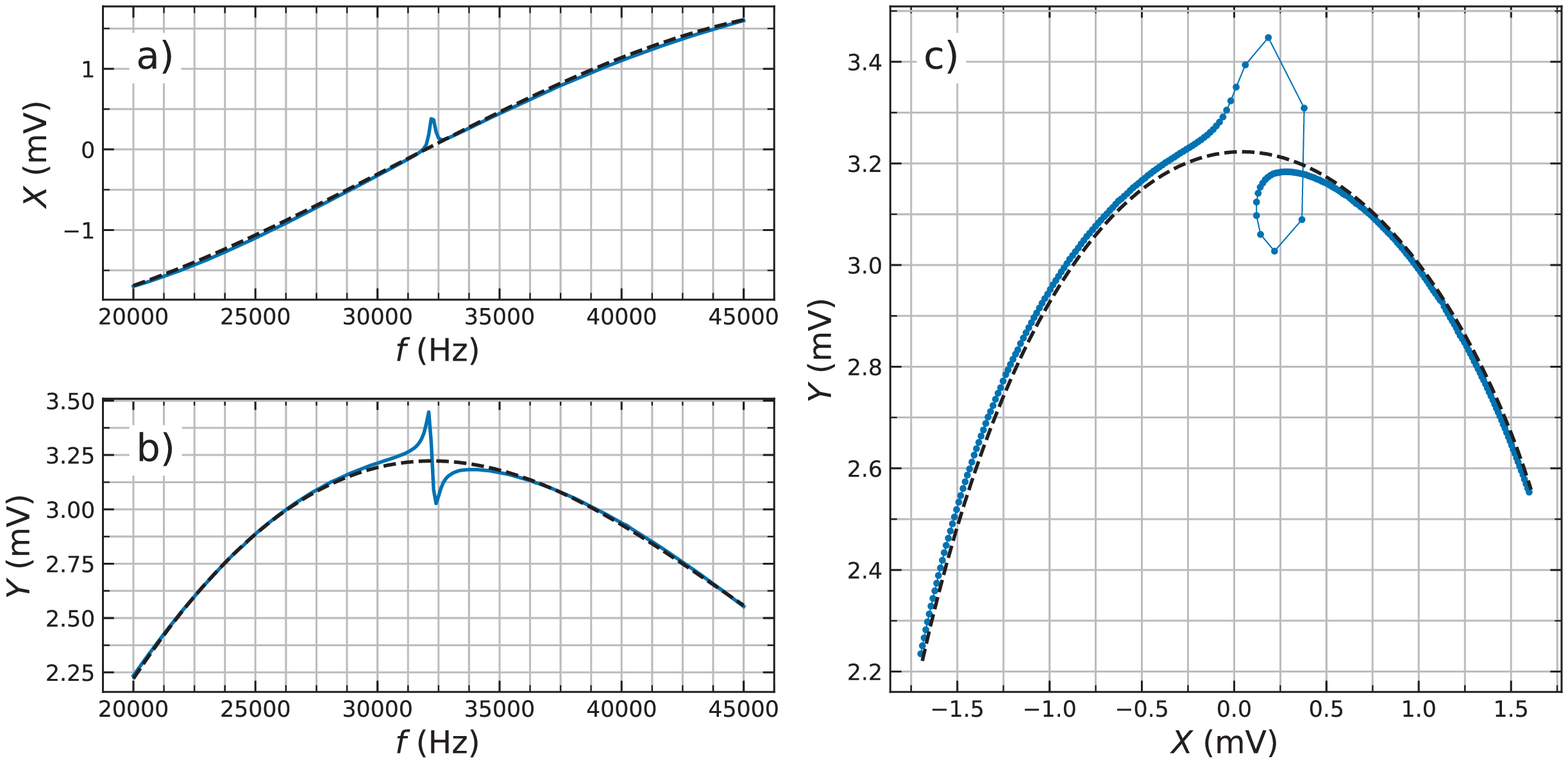}
\caption{
    {\bf{Quartz tuning fork background sweep}.} The response of a quartz tuning fork at 8.1 mK and 29.3 bar over a wide frequency range.  
    The dashed lines are the fitted 3$^{\rm{rd}}$ order and 4$^{\rm{th}}$ order polynomials for the $X$ a) and $Y$ b) responses after the correction noted to fit a Lorentzian to the fork's resonance (see Supplemental Figure~\ref{fig::S4}). In panel c) we compare the fit (dashed line) to the observed complex response. }
\label{fig::S3}
\end{figure}

To effect this subtraction, we first carried out a sweep from 20 kHz through 45 kHz at $\approx$ 100 mK (not shown here) where the resonant line is narrow. This was done to assess the background and allow the fork to be operated while cooling down to dilution refrigerator temperatures. At lower temperatures (8.1 mK shown in Supplementary Figure~\ref{fig::S3}), we repeated this sweep. We then swept the frequency through the resonance over a frequency range spanning a few linewidths.
We found that the $X$ channel could be fit to a 3$^{\rm{rd}}$ order polynomial, and the $Y$ channel required a 4$^{\rm{th}}$ order polynomial to adequately fit the background. After subtraction of these backgrounds, we plot the narrow range signal and obtain a Lorentzian fit for the resonance. 
A Nyquist plot with the real axis aligned along the $X$ channel and the out-of-phase response aligned along $Y$ reveal a near perfect circle plot. These $X$ and $Y$ signals (after subtraction of the fitted background), together with the associated Nyquist plot are shown in Supplementary Figure 4. In order to obtain a satisfactory Lorentzian fit a small ($\le$ +0.05 mV) shift to the fitted background was needed. This increase in $X$ accounts for the difference between the fit used (shown as a dashed line) and the data shown in Supplementary Figure 3. The Lorentzian is used to obtain the $Q = Q_R$, and the amplitude of the signal at resonance ($A_R$) is also noted. The previously mentioned constant $k$ is defined as $k = Q_R/A_R$. (The value of $k$ used in the LabView VI was likely not accurate enough and necessitated adjustments described in the following sections). Together with the $X(T)$ and $Y(T)$ ($X, Y$ values after subtraction of the background at any temperature $T$), these constitute the inputs to the determination of the resonant frequency $f(T)$ and the $Q(T)$ using the equations\cite{Morley2002}, 

\begin{equation}\label{supeq1}
Q(T) = \frac{X(T)^2+Y(T)^2}{X(T)}\left(\frac{Q_R}{A_R}\right)
\end{equation}

\begin{equation}\label{supeq2}
f(T) = f_D(T)\left(1+\frac{Y(T)}{X(T)}\frac{1}{2Q_T}\right)
\end{equation}

\noindent where $f_D(T)$ is the drive frequency at any temperature, $T$. These equations form the basis of the PLL that maintains the fork within $\pm$ 5 Hz of resonance and were used to calculate the ``raw" values of $Q(T)$ and $f(T)$.

 \begin{figure}[H]
 \renewcommand{\figurename}{Supplementary Figure}
\centering
\includegraphics[width=\linewidth, keepaspectratio]{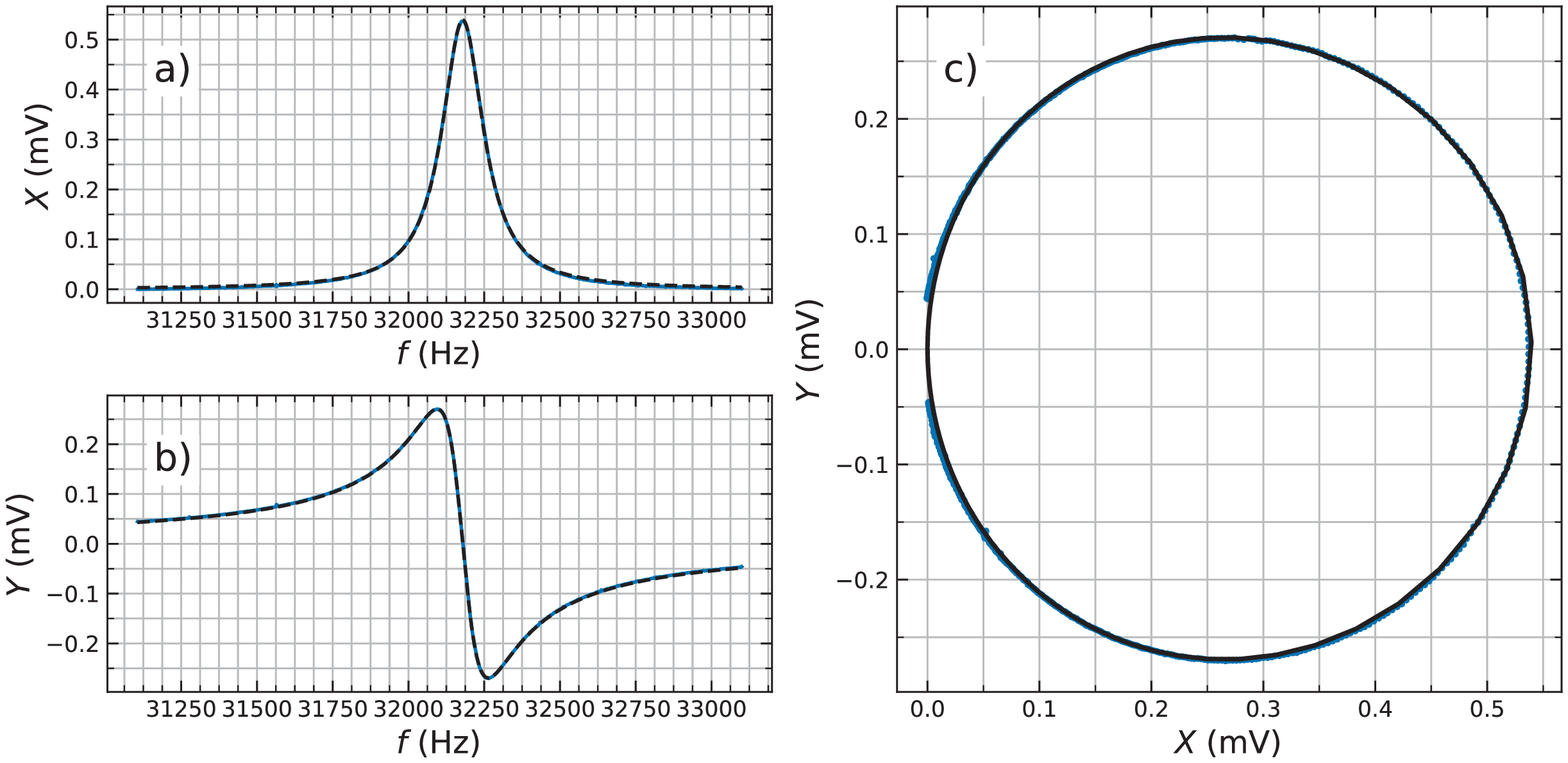}
\caption{{\bf{Quartz tuning fork Lorentzian response}.} The response of a quartz tuning fork at 8.1 mK and 29.3 bar after subtraction of the background.  The black (dashed/solid) lines on the (left/right) are the fitted curves to a Lorentzian response.}
\label{fig::S4}
\end{figure}

When we apply the same background subtraction to a frequency sweep at 4.9 mK (obtained a few days later), where the $Q$ is further reduced, the Nyquist response within the linewidth is horizontally off-center. This offset in the $X$ background is not systematically temperature dependent. Instead it appears that there is a small frequency dependence to the background (corresponding to a first order term) that is not captured in our background subtraction procedure.

 \begin{figure}[H]
 \renewcommand{\figurename}{Supplementary Figure}
\centering
\includegraphics[width=\linewidth, keepaspectratio]{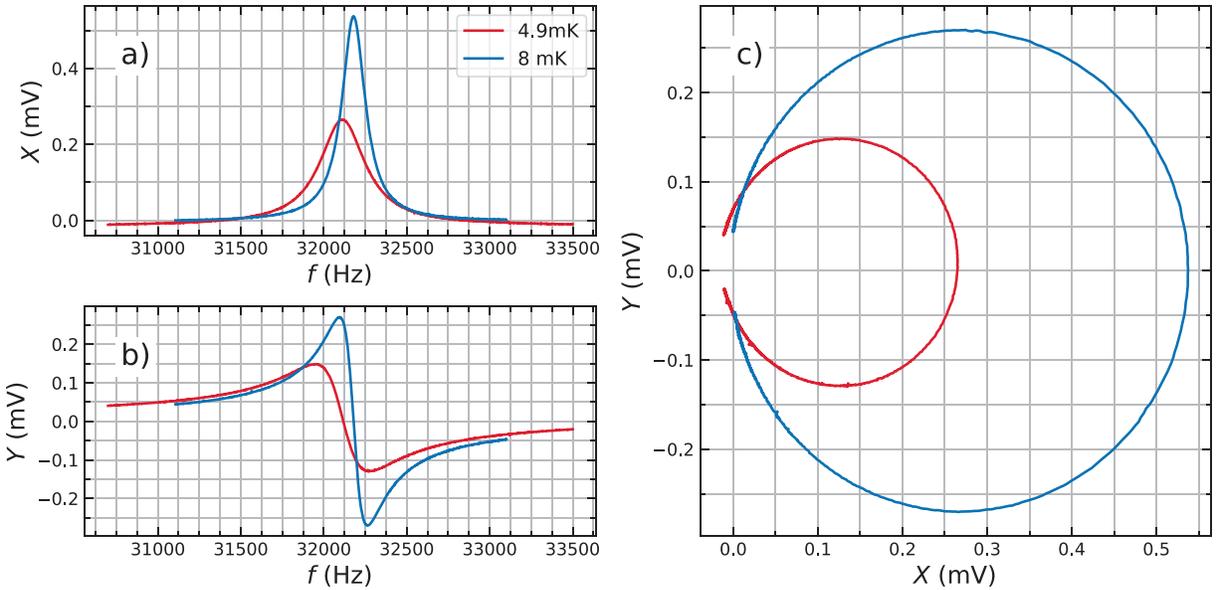}
\caption{
    {\bf{Quartz tuning fork Lorentzian responses at 8.1 mK and 4.9 mK}.} 
    The response of a quartz tuning fork at 29.3 bar taken at 8.1 mK (blue) compared to the response at 4.9 mK (red), both obtained after subtraction of the fitted background signal obtained at 8.1 mK (See Supplementary Figure~\ref{fig::S4}c)).  The Nyquist plot for the 4.9 mK signal shows an offset from the origin. This is also visible as a negative response for the $X$ signal at 4.9 mK (red) in a).
    }
\label{fig::S5}
\end{figure}

\begin{figure}[H]
 \renewcommand{\figurename}{Supplementary Figure}
\centering
\includegraphics[width=0.75\linewidth, keepaspectratio]{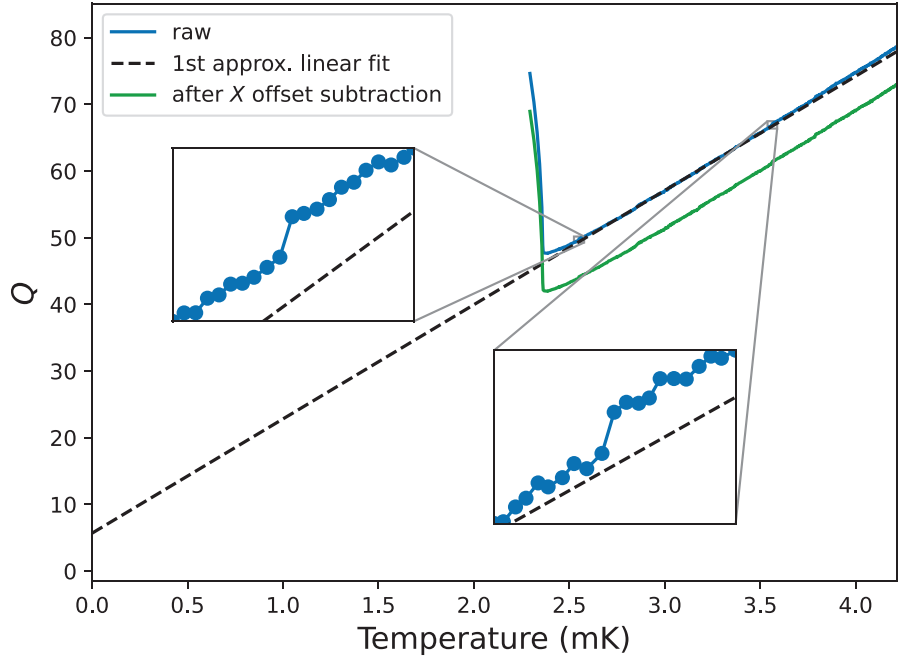}
\caption{
    {\bf{$Q$ vs T before and after offsets at $P = 29.3$ bar}.} 
    Temperature dependence of $Q$ (blue line) inferred following subtraction of the non-resonant signal background. 
    The dashed line is a linear fit on a $10$ percent random sample between $1.2T_c$ and $1.4T_c$.
    The temperature fitting range was kept narrow in order to avoid fitting jumps in the data.
    After subtraction of the offset, we plot the resulting $Q$ vs $T$ in green. }
\label{fig::S6}
\end{figure}

  Because the origin of this shift in background was not evident when the data was being acquired, we introduced a workaround to compensate for this shift.  We collect data in a limited period of time (typically 1 day) while ensuring minimal thermal gradient between the thermometer immersed in the $^3$He and the quartz fork. In practice, this limits us to temperature sweeps within $\approx$ 40\% of $T_c$. Following this procedure, we observe an artificial intercept ($Q \neq 0$ at $T$ = 0) in the temperature dependence of the $Q$ in $^3$He (Supplementary Figure~\ref{fig::S6}). After subtraction of this intercept ($Q(0)_1$), the inferred $Q$ is plotted as the green line in Supplementary Figure~\ref{fig::S6}. Intercepts of this magnitude or smaller were observed for all the different pressure runs. A list of the offset $Q(0)_1$ values subtracted for each pressure is shown in Supplementary Table~\ref{tab:Qadjtable}.

As the $^3$He is cooled, due to the increased viscosity, the mass of $^3$He coupled to the fork changes, resulting in a decrease of the resonant frequency. Since the PLL operates at a fixed drive frequency, we reset the drive frequency once $f_D$ -$f_0(T) \ge$ 5 Hz.  However, because of the offset in $X$ that shifted the $Q$ intercept, Equations~\ref{supeq1}, \ref{supeq2} are no longer exactly valid.  This leads to artificial jumps in the inferred $Q$ produced when the drive frequency is reset (see the insets to Supplementary Figure~\ref{fig::S6}).  We note that it was difficult to calculate a line of best fit for linear data broken up into slightly offset line segments as seen in Supplementary Figure~\ref{fig::S6}. This was remedied after the $Q$ continuity correction.


A line was fitted to the $Q(T)$ data, and the $Q(0)_1$ value was converted to a voltage, $X_{\rm offset}$. The $Q(0)_1$ values vary between $-1 < Q < 6$ with pressure and  are listed in Supplementary Table~\ref{tab:Qadjtable}. After subtracting the $X_{\rm offset}$ from the raw fork response, we carry out the corrections to achieve the continuity of $Q$ as described next.

In Supplementary Figure~\ref{fig::S7} (a), we plot the values of $X(T)$, $Y(T)$ obtained during a cooldown at 29.3 bar.  
The traces are broken up into segments of data collected at a fixed drive frequency. 
At each transition the drive frequency was reset by the PLL to be on resonance ($Y(T)$ = 0). 
In the top panel of Supplementary Figure~\ref{fig::S7}, at each point where a frequency reset is triggered, a circular arc is traced. 
This arc of constant $Q$ corresponds to a segment of a Nyquist plot for a circle of diameter corresponding to the inferred $Q$ factor just before the reset, centered at $(X,Y)$ corresponding to $Q$/2, 0. 
After the frequency reset, the green trace corresponding to the uncorrected data, systematically deviates away from the arc. 
This indicates that the background fit used has a small systematic frequency dependent error that was not resolved in the fitting procedure described earlier to obtain the Lorentzian fit shown in Supplementary Figure~\ref{fig::S4}. 
To correct for this offset, the next segment of the data is shifted to the preceding arc of constant $Q$. 
For the specific case of the 29.3 bar data shown here, each segment was shifted by a positive increment to $X$, corresponding to a positive shift in $Q$. 
We find that the individual changes to $\overline{\Delta Q_1}$ are of order $0.2 \pm 0.1$.
The resulting corrected trace is shown in red in the bottom panel of Supplementary Figure~\ref{fig::S7}. This concludes the first iterative correction. 

 \begin{figure}[H]
 \renewcommand{\figurename}{Supplementary Figure}
\centering
\includegraphics[width=\linewidth, keepaspectratio]{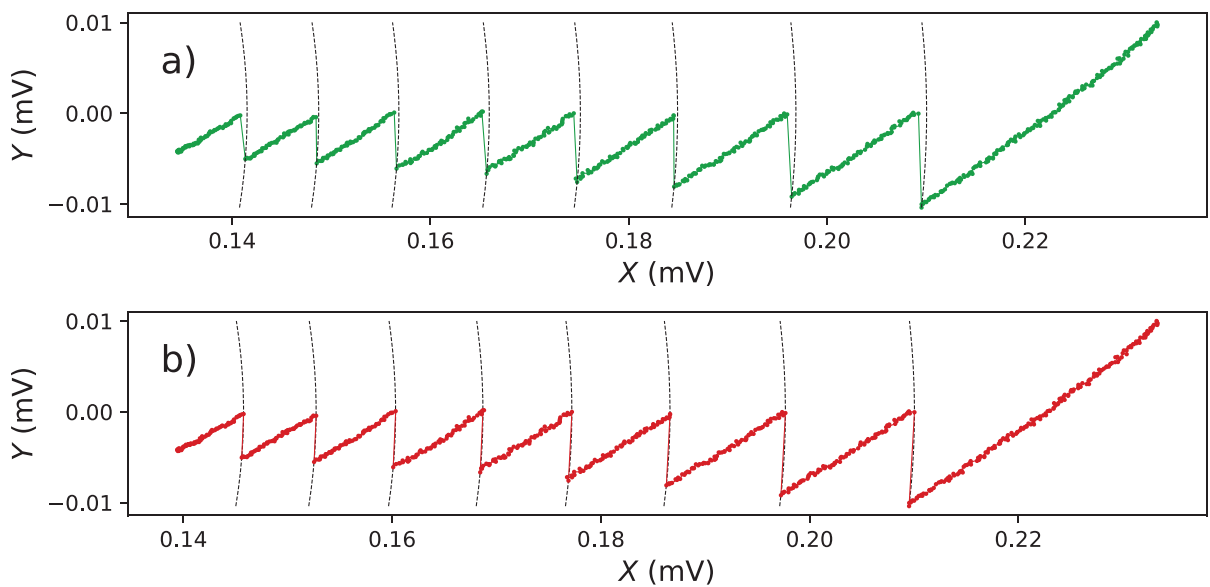}
\caption{
    {\bf{Nyquist trace during a cooldown at $P = 29.3$ bar}.} 
    The top trace shows the observed trace of $Y(T)$ vs $X(T)$ data taken while cooling before correction (green). As the temperature decreases, the $X(T)$ decreases, and $Y(T)$ also decreases corresponding to a decrease in the resonant frequency. Also shown are individual segments of circles (Nyquist plots) originating at $X$ = 0, $Y$ = 0 that pass through the last point in $X(T)$, $Y(T)$ before the frequency is reset. The lower panel depicts the trace (red) of $X(T)$ vs $Y(T)$ after adjustment described in the text.}
\label{fig::S7}
\end{figure}

\section*{Second and Third Iterative Correction.}

Because of the cumulative change in the $Q$ accompanying each reset of the drive frequency, the $T$ = 0 intercept seen in Supplementary Figure~\ref{fig::S3} would be changed. Therefore, a further two iterations to obtain the offsets $Q(0)_2$ and $Q(0)_3$  and $Q$ continuity corrections yielding $\overline{\Delta Q_2}, \overline{\Delta Q_3}$ were carried out to minimize the $T=0$ intercept and any remaining discontinuities in $Q$ across resets of drive frequency. These linear fits were extended to between 1.2 $T_c$ and 2$T_c$  (or the highest temperature that data at a given pressure was acquired at).  The final corrected data and linear fit is shown as the color coded trace in Figure 1 of the main paper, and was used in all further analysis detailed in the main paper. The values for $Q(0)_2$, $Q(0)_3$ and $\overline{\Delta Q_2}, \overline{\Delta Q_3}$ are listed in Supplementary Table~\ref{tab:Qadjtable}. 

\section*{Correction to $k$.}

The frequency dependent correction described in the previous sections is essentially confined to $X(T)$. Consequently, the raw inferred resonance frequency $f_0(T)$ is continuous with temperature (blue trace in Supplementary Figure~\ref{fig::S8}). 
However, after making corrections to the $X$ component of the response as detailed in this Supplementary Note, the inferred frequency is no longer continuous across changes in drive frequency (red trace in Supplementary Figure~\ref{fig::S8}). A correction is needed to $k$ (the difference in the value of the inferred frequency from the drive frequency depends on the value of $k$ - See Supplementary Equations~\ref{supeq1}, \ref{supeq2}). 
Note that $Y(T) \ll$ $X(T)$, so that $Q(T) \approx X(T)\times k$. Therefore, we apply a multiplicative constant $k_{adj} = k'/k$, where $k'$ is the new voltage to $Q$ conversion factor. 
The $k_{adj}$ is found by minimizing the sum of all the jumps at a frequency reset, resulting in the dashed orange trace in Supplementary Figure~\ref{fig::S8} that aligns well the raw resonance frequency.  
In a few runs (8 bar, 5 bar and 2 bar) the drive frequency was held fixed throughout the temperature sweep. Consequently, there are no values for $\overline{\Delta Q_1}, \overline{\Delta Q_2}, \overline{\Delta Q_3}$ in the table. In these runs, the zero temperature offset was subtracted, and the $k_{adj}$ was found by minimizing the difference in the original raw $f_o$ and the $f_o$ upon a $X$ offset subtraction based on a zero temperature intercept. 

 \begin{figure}[H]
 \renewcommand{\figurename}{Supplementary Figure}
\centering
\includegraphics[width=0.6\linewidth, keepaspectratio]{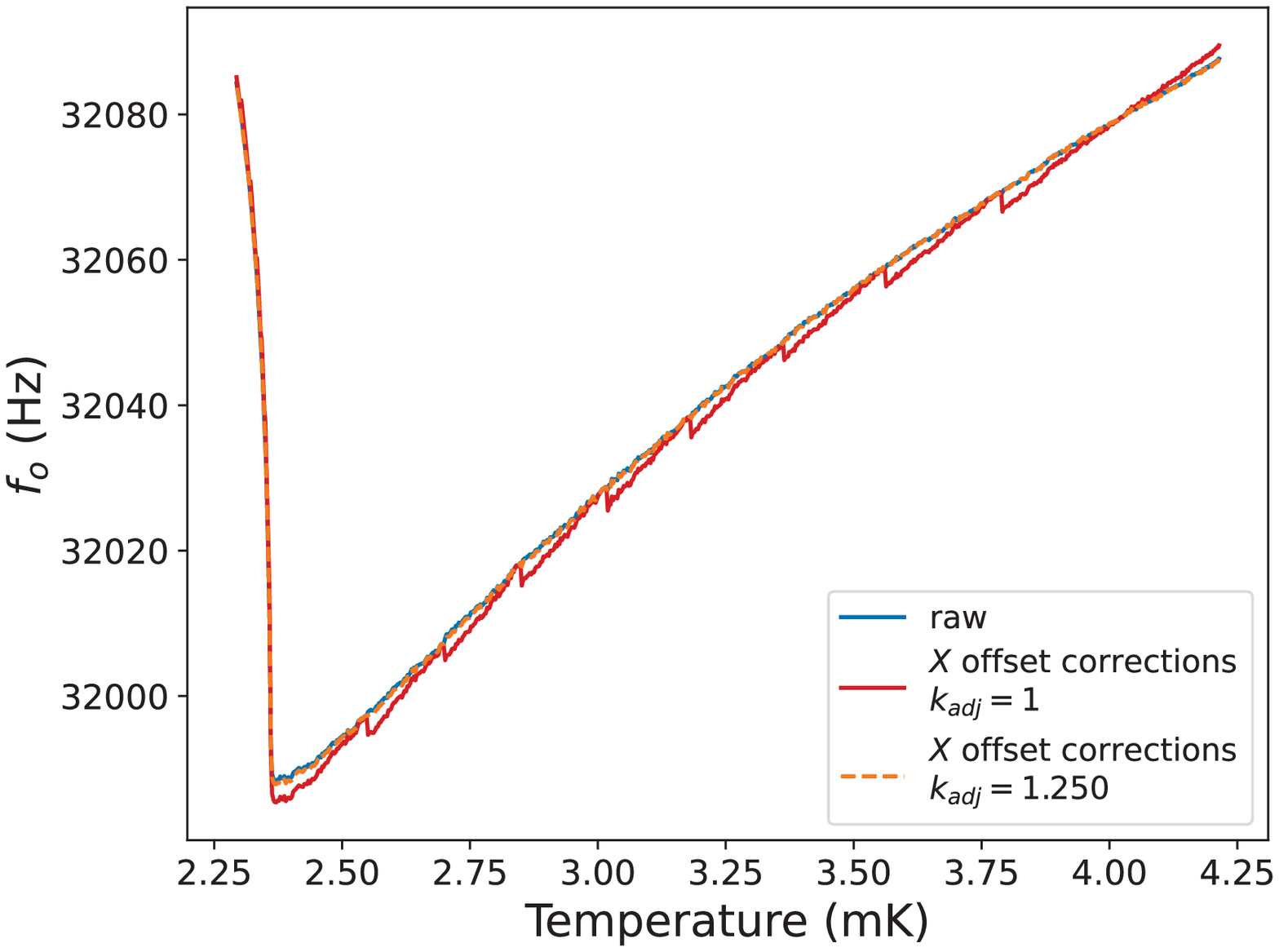}
\caption{
    {\bf{Resonant Frequency {\it vs} Temperature at $P=29.3$ bar}.} 
   The blue trace shows the raw data for resonant frequency plotted against the temperature, obtained from the fit to $Q$ at 8.1 mK. Following offsets of $Q$, the resulting plot (orange) displays discontinuities. This is resolved by appropriate changes to $k_{adj}$ restoring $f_0(T)$ (dashed red line) to the original values.  }
\label{fig::S8}
\end{figure}

After following these steps, we plot the corrected results for $Q$ vs $T$ and $\delta Q$ vs T alongside the uncorrected raw data in Supplementary Figures~\ref{fig::S9}, \ref{fig::S10}. We also show the full extent of all temperatures and pressures measured with corrections applied in Supplementary Figure~\ref{fig::S11}. Individual plots similar to Supplementary Figure~\ref{fig::S9} for 27 bar, 23 bar, 20 bar and 15 bar are shown in Supplementary Figure~\ref{fig::S12} and for 8 bar, 5 bar, 2 bar and 0.5 bar are shown in Supplementary Figure~\ref{fig::S13}.

 \begin{figure}[H]
 \renewcommand{\figurename}{Supplementary Figure}
\centering
\includegraphics[width=0.97\linewidth, keepaspectratio]{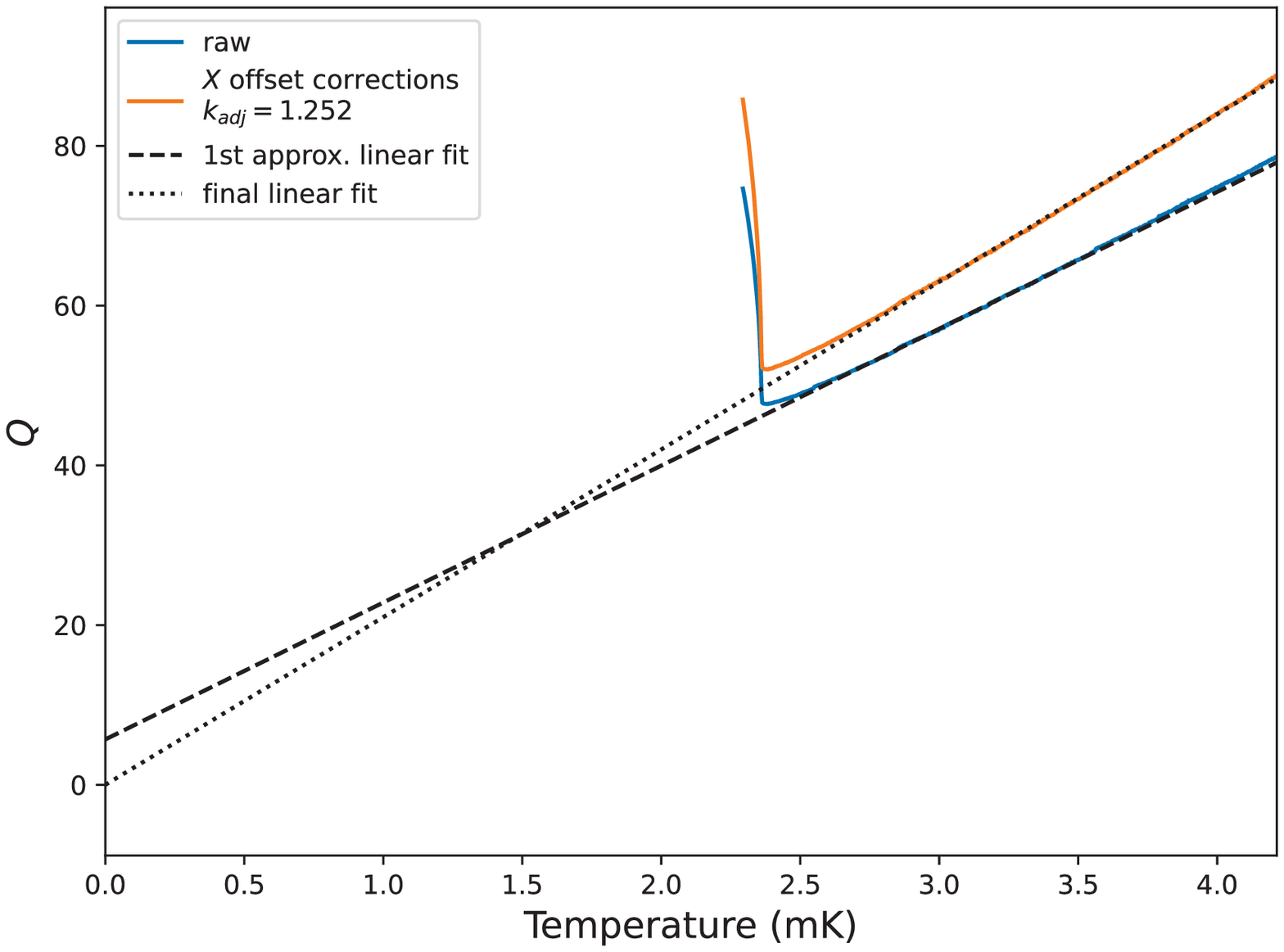}
\caption{
{\bf{\it Q vs.} Temperature before and after $X$ offset corrections $P=29.3$ bar}. 
    We show the $Q$ before (blue) and after (orange) the $X$ offset and $k$ correction were applied. 
    The plot in shown in gold here appears in Fig 1 a) of the main paper.
    The dashed line is a linear fit on the raw data in the 1st iteration of the correction procedure.
    The dotted line is a linear fit on the fully corrected data.}
\label{fig::S9}
\end{figure}

 \begin{figure}[H]
 \renewcommand{\figurename}{Supplementary Figure}
\centering
\includegraphics[width=\linewidth, keepaspectratio]{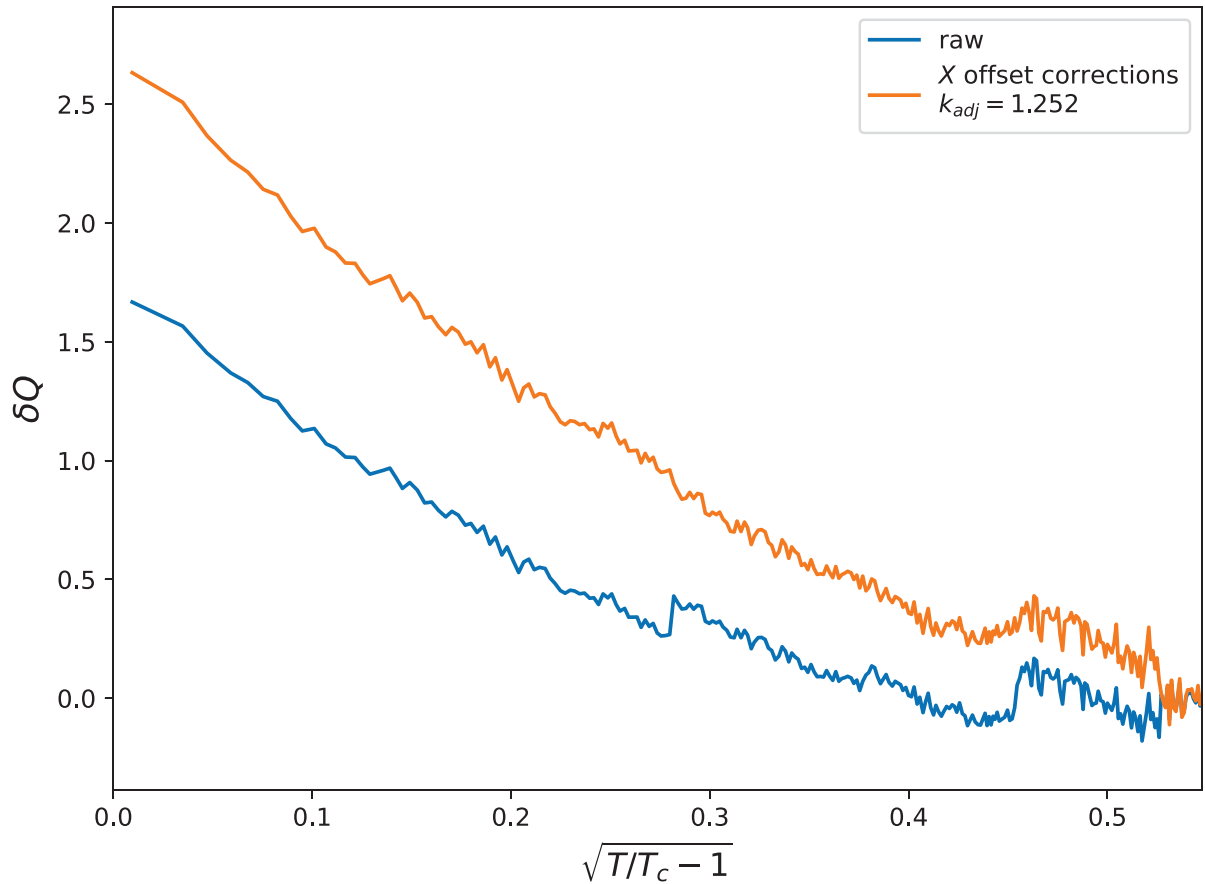}
\caption{
    {\bf{
    {\it $\delta$Q vs.}} the square root of the reduced temperature before and after offsets at $P=29.3$ bar.
    }
   We show the $\delta Q$ before (blue) and after (gold) the correction offsets were applied. The plot in shown in gold here appears in Figure 2a) of the main paper.}
\label{fig::S10}
\end{figure}

 \begin{figure}[H]
 \renewcommand{\figurename}{Supplementary Figure}
\centering
\includegraphics[width=\linewidth, keepaspectratio]{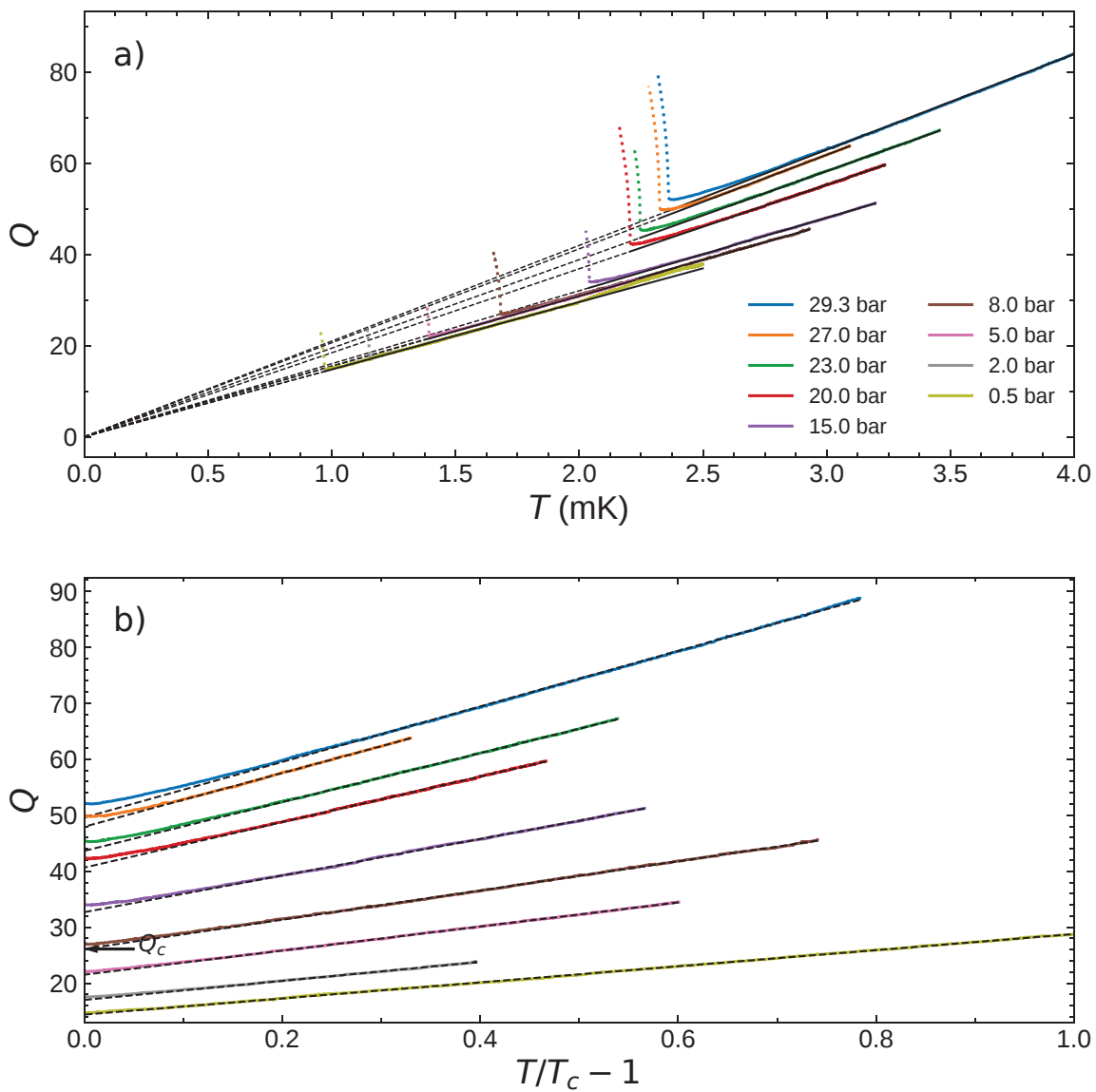}
\caption{
    {\bf{Corrected $Q$ vs. temperature of all pressures over a larger temperature range.}
    }
    We show the corrected $Q$ vs. temperature data up to the highest temperature data was taken for each pressure. 
    a) Corrected $Q$ vs temperature, where the dashed lines are linear fits.
    The dotted colored line is the fork response below $T_c$.
    The dashed black lines are the linear fits extrapolated below $T_c$.
    b) $Q$ versus a reduced temperature scale.}
\label{fig::S11}
\end{figure}

 \begin{figure}[H]
 \renewcommand{\figurename}{Supplementary Figure}
\centering
\includegraphics[width=\linewidth, keepaspectratio]{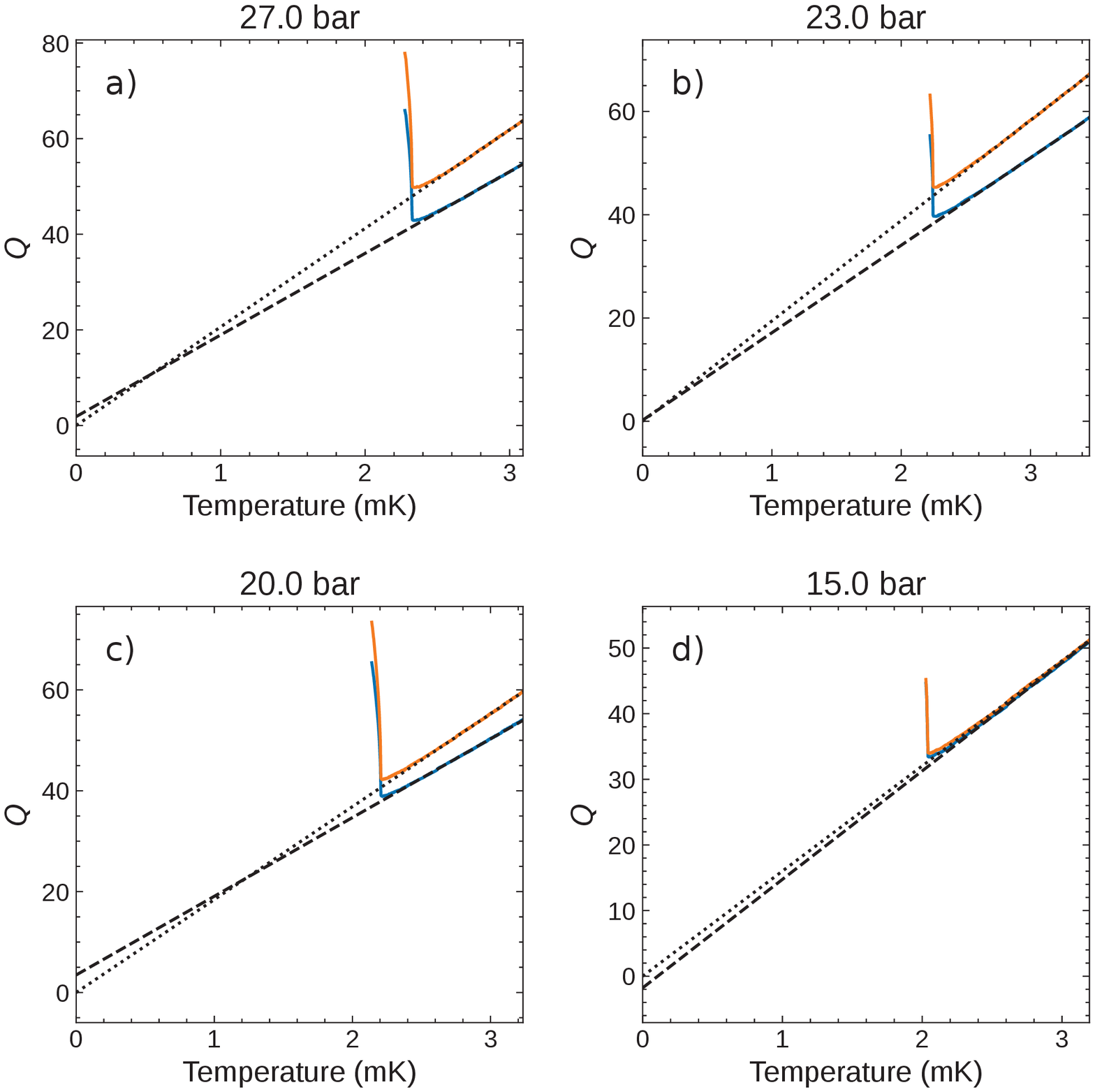}
\caption{
    {\bf{Corrected $Q$ vs. temperature for a) 27 bar, b) 23 bar, c) 20 bar, and d) 15 bar.}
    }
    The legend for the traces plotted in each section of the grid are equivalent to Supplementary Figure~\ref{fig::S9}.
    The plot compares the corrected data (orange), to the ``as collected" data (blue).
    The $k_{adj}$ constants for each run can be found in table~\ref{tab:Qadjtable}.
    The dashed line is a linear fit to the uncorrected data, and the dotted line is the final linear fit to the corrected data.
    }
\label{fig::S12}
\end{figure}

 \begin{figure}[H]
 \renewcommand{\figurename}{Supplementary Figure}
\centering
\includegraphics[width=\linewidth, keepaspectratio]{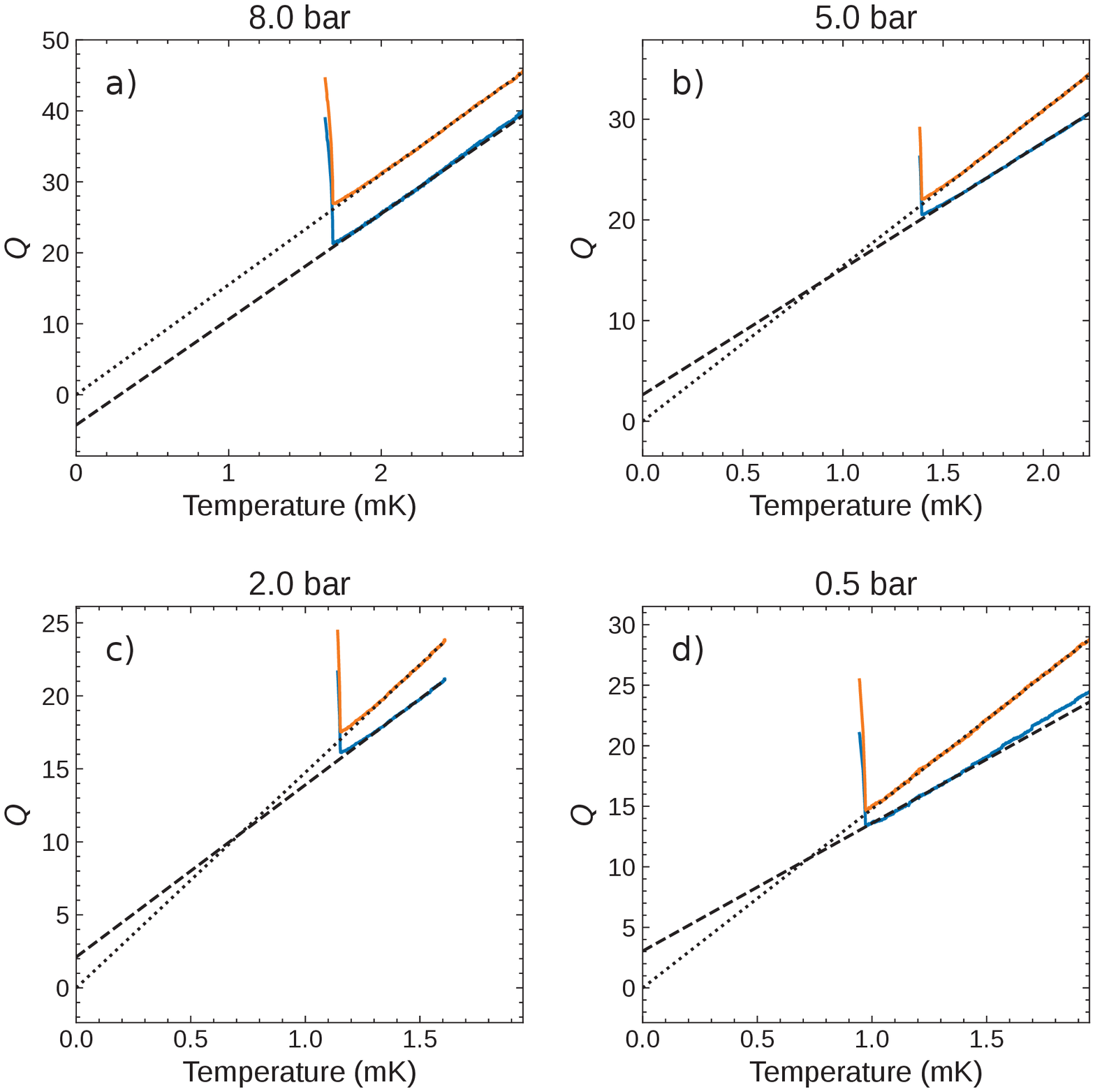}
\caption{
    {\bf{Corrected $Q$ vs. temperature for a) 8 bar, b) 5 bar, c) 2 bar, and d) 0.5 bar.}
    }
    The legend for the traces plotted in each section of the grid are equivalent to Supplementary Figure~\ref{fig::S9}.
    The plot compares the corrected data (orange), to the``as collected" data (blue).
    The $k_{adj}$ constants for each run can be found in table~\ref{tab:Qadjtable}.
    The dashed line is a linear fit to the uncorrected data, and the dotted line is the final linear fit to the corrected data.
    }
\label{fig::S13}
\end{figure}

\pagebreak
\begin{sidewaystable}
    \centering
    \begin{tabular}{ |c|c|c|c|c|c|c|c|c| }  
    \hline
     P (bar) & final $Q(0)$ & $k_{adj}$ & $Q(0)_1$ & $Q(0)_2$ & $Q(0)_3$ & 1st $\overline{\Delta Q_1}$ & $\overline{\Delta Q_2}$ & $\overline{\Delta Q_3}$ \\ 
    \hline
    29.3	 &   0.00252	  &  1.25	&    5.69	&    1.95	    &    0.059	     &   -0.231	    &    -0.0025	 &   -0.000103  \\
    27	 &   4.13e-05     &  1.22	&    1.86	&    0.594      &    0.00431	 &   -0.00353	&    -9.13e-05	 &   -7.53e-07	  \\
    23	 &   -1.3e-05	  &  1.14	&    0.208	&    -0.182	    &    -0.00154	 &   -0.00112	&    3.32e-05	 &   2.23e-07   \\
    20	 &   -3.81e-05	  &  1.18	&    3.48	&    -0.0172	&    -0.00146	 &   -0.0829	&    0.00019	 &   1.35e-06   \\
    15	 &   0.000169	  &  1	    &    -1.78	&    1.47	    &    0.0157	     &   -0.024	    &    -0.000407   &	-4.22e-06   \\
    8	 &   -1.03e-07	  &  1	    &    -4.26	&    -1.36	    &    9.78e-06	 &   0          &    0           &   0  \\
    5	 &   0.000111	  &  1.23	&    2.63	&    0.0485	    &    -0.00209	 &   0	        &    0	         &   0  \\
    2	 &   -5.64e-05	  &  1.25	&    2.12	&    -0.0543	&    0.00154	 &   0	        &    0	         &   0  \\
    0.5	 &   2.64e-05	  &  1.42	&    3.04	&    0.0837	    &    -0.00319	 &   0.0696	    &    -0.00018	 &   9.17e-07   \\
     \hline
    \end{tabular}
     \caption{\label{tab:Qadjtable} {\bf Supplementary Table 2: Background Subtraction.} 
     The table lists the pressures, the final value of $Q(T=0)$, the adjusted value of $k$ (the multiplicative factor applied the conversion factor $k$ from signal amplitude to $Q$), the $Q(T=0)$ for three iterations of the correction procedure, and the the average jump in $Q$ after a frequency reset for three iterations.}
\end{sidewaystable}

\pagebreak
\clearpage

\section*{Background in the low-$Q$ regime}

\begin{figure}[H]
 \renewcommand{\figurename}{Supplementary Figure}
\centering
    \includegraphics[width = 1\textwidth]{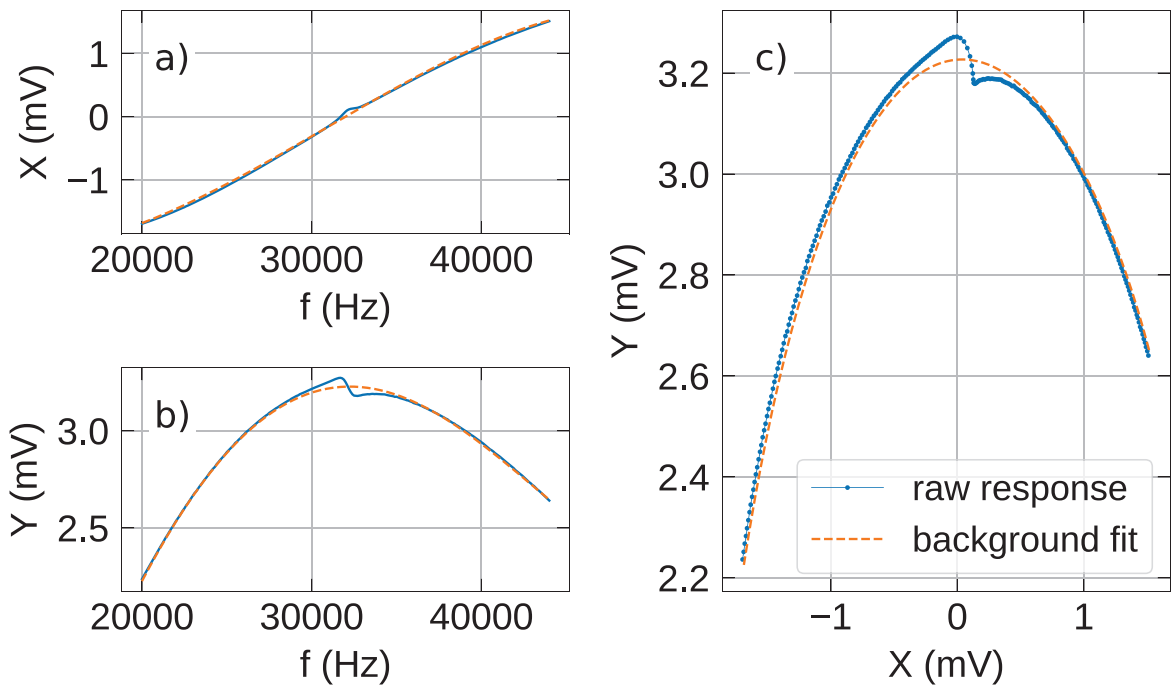}
    \caption{
        \textbf{Background subtraction over a wide frequency range in the low $Q$ regime.}
        $T=2.5$ mK and $P=0.56$ bar.  
        a) The $X$ response of the quartz fork fitted to a 3$^\text{rd}$ order polynomial.
        b) The $Y$ response of the quartz fork fitted to a 4$^\text{th}$ order polynomial.
        c) The fork response over a wide frequency range in a Nyquist plot, together with the fit (orange dashed line).
    }
    \label{fig::Sfig14}
\end{figure}

\begin{figure}[H]
 \renewcommand{\figurename}{Supplementary Figure}
\centering
    \includegraphics[width = 1\textwidth]{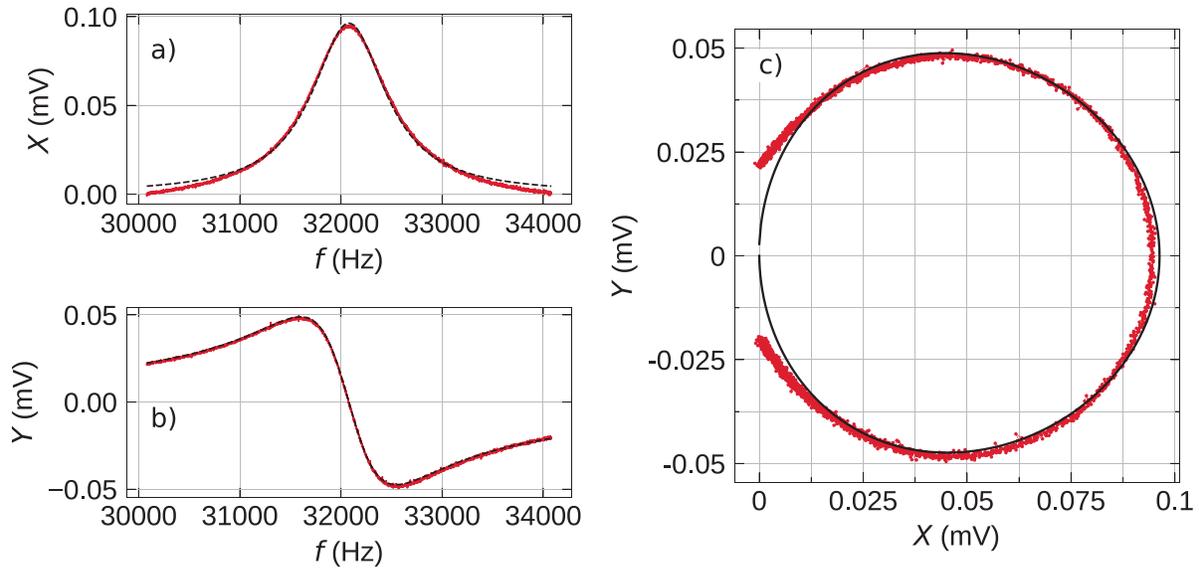}
    \caption{
        \textbf{Background subtraction over a narrow frequency range in the low $Q$ regime.}
        $T=2.5$ mK and $P=0.56$ bar. 
        The $Q$-factor from this fit is $\approx 35$, and the fitted resonance is $f_0=32079$ Hz (linewidth, $\Delta f=916$ Hz).
        a) The $X$ response of the quartz fork fitted to the real part of a complex Lorentzian.
        b) The $Y$ response of the quartz fork fitted to the imaginary part of a complex Lorentzian.
        c) The quartz fork response and fit in a Nyquist plot.  }
            \label{fig::Sfig15}
\end{figure}

In addition to the background presented in Supplementary Figures~\ref{fig::S3},~\ref{fig::S4}, we carried out a frequency sweep to fit the background at low temperature and low pressure.
Supplementary Figure~\ref{fig::Sfig15}a) shows poor agreement between the real response and the fit to a Lorentzian.  
The fit deviation in the $X$ component is observed in the Nyquist plot in Supplementary Figure~\ref{fig::Sfig15} a) and c) as well.
The fork's response is small in the low-$Q$ regime. The broader response due to the low $Q$ and a temperature dependent background, is responsible for a poor fit. In Supplementary Figure~\ref{fig::Sfig16} we show the consequence of using a high temperature and high pressure fit to data obtained in the low $Q$ regime.   

\begin{figure}[H]
  \renewcommand{\figurename}{Supplementary Figure}
 \centering
     \includegraphics[width = 1\textwidth]{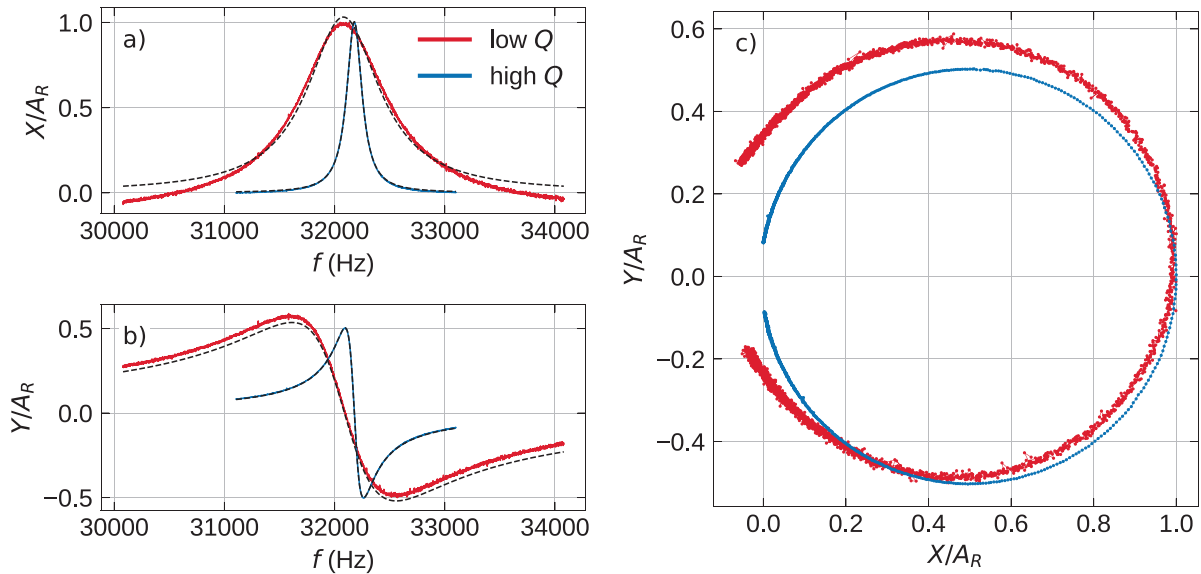}
     \caption{
         \textbf{Low $Q$ sweep with a high $Q$/high temperature and pressure background.}
         The blue trace is the response from Supplementary figure~\ref{fig::S4} minus the background obtained from the wide response in figure~\ref{fig::S3}, at $T=8.1$ mK and P = $29.3$ bar.
         The red trace is the fork response at $T=2.5$ mK, $P=0.56$ bar, minus the same high temperature and high pressure background as the blue trace. 
         a) The $X$ responses of the quartz fork fitted to the real part of a complex Lorentzian, and normalized by the resonant amplitude.
         b) The $Y$ responses of the quartz fork fitted to the imaginary part of a complex Lorentzian and normalized by the resonant amplitude.
         c) The quartz fork responses in a Nyquist plot normalized by the resonant amplitude. }
             \label{fig::Sfig16}
 \end{figure}

\pagebreak
\clearpage


\providecommand{\noopsort}[1]{}\providecommand{\singleletter}[1]{#1}%


\end{document}